\documentclass[a4paper,fleqn,usenatbib]{mnras}

\usepackage{times}

\usepackage[T1]{fontenc}
\usepackage{ae,aecompl}

\usepackage{graphicx}	
\usepackage{amsmath}	
\usepackage{amssymb}	

\defcitealias{Sobral2015}{S15}
\defcitealias{Matthee2017_ALMA}{M17}
\newcommand\gsim{~\lower.5ex\hbox{$\buildrel > \over \sim$}~}
\newcommand\lsim{~\lower.5ex\hbox{$\buildrel < \over \sim$}~}

\title[On the resolved nature of CR7]{ On the nature and physical conditions of the luminous Ly$\alpha$ emitter CR7 and its rest-frame UV components\thanks{Based on observations obtained with {\it HST}/WFC3 {program 14495} and the VLT programs 294.A-5018 and 294.A-5039.}}

\author[D. Sobral et al.]{David Sobral$^{1,2}$\thanks{E-mail: d.sobral@lancaster.ac.uk}, Jorryt Matthee$^{2}$, Gabriel Brammer$^{3}$, Andrea Ferrara$^{4,5}$, Lara Alegre$^{6}$, \newauthor Huub  R\"{o}ttgering$^{2}$, Daniel Schaerer$^{7,8}$, Bahram Mobasher$^{9}$, Behnam Darvish$^{10}$ \\
$^{1}$ Department of Physics, Lancaster University, Lancaster, LA1 4YB, UK \\
$^{2}$ Leiden Observatory, Leiden University, P.O.\ Box 9513, NL-2300 RA Leiden, The Netherlands \\
$^{3}$ Space Telescope Science Institute, 3700 San Martin Dr, Baltimore MD 21211, USA \\
$^{4}$ Scuola Normale Superiore, Piazza dei Cavalieri 7, I-56126 Pisa, Italy \\
$^{5}$ Kavli IPMU, The University of Tokyo, 5-1-5 Kashiwanoha, Kashiwa 277-8583, Japan \\
$^{6}$ Institute for Astronomy, University of Edinburgh, Royal Observatory, Blackford Hill, Edinburgh EH9 3HJ, UK \\
$^{7}$ Observatoire de Gen\`eve, Universit\`e de Gen\`eve, 51 Ch. des Maillettes, 1290 Versoix, Switzerland \\
$^{8}$ CNRS, IRAP, 14 Avenue E. Belin, 31400 Toulouse, France \\
$^{9}$ Department of Physics and Astronomy, University of California, 900 University Ave., Riverside, CA 92521, USA \\
$^{10}$ Cahill Center for Astrophysics, California Institute of Technology, 1216 East California Boulevard, Pasadena, CA 91125, USA}

\date{Accepted 2018 October 10. Received 2018 October 03; in original form 2017 October 21}

\pubyear{{2018}}

\begin{document}
\label{firstpage}
\pagerange{\pageref{firstpage}--\pageref{lastpage}}
\maketitle

\begin{abstract}

\noindent We present new {\it HST}/WFC3 observations and re-analyse VLT data to unveil the continuum, variability and rest-frame UV lines of the multiple UV clumps of the most luminous Ly$\alpha$ emitter at $z=6.6$, CR7. Our re-reduced, flux calibrated X-SHOOTER spectra of CR7 reveal a He{\sc ii} emission line in observations obtained along the major axis of Ly$\alpha$ emission with the best seeing conditions. He{\sc ii} is spatially offset by $\approx+0.8''$ from the peak of Ly$\alpha$ emission, and it is found towards clump B. Our WFC3 grism spectra detects the UV continuum of CR7's clump A, yielding a power law with {$\beta=-2.5^{+0.6}_{-0.7}$ and $M_{UV}=-21.87^{+0.25}_{-0.20}$}. {No significant variability is found for any of the UV clumps on their own, but there is tentative ($\approx2.2$\,$\sigma$) brightening of CR7 in F110W as a whole from 2012 to 2017. {\it HST} grism data fail to robustly detect rest-frame UV lines in any of the clumps, implying fluxes$\lsim2\times10^{-17}$\,erg\,s$^{-1}$\,cm$^{-2}$ (3\,$\sigma$). We perform {\sc cloudy} modelling to constrain the metallicity and the ionising nature of CR7}. CR7 seems to be actively forming stars without any clear AGN activity in {clump A, consistent with a metallicity of $\sim0.05-0.2$\,Z$_{\odot}$. Component C or an inter-clump component between B and C may host a high ionisation source}. Our results highlight the need for spatially resolved information to study the formation and assembly of early galaxies.
 
\end{abstract}


\section{Introduction}

The significant progress in identifying large samples of distant galaxies \citep[e.g.][]{Bouwens2015,Harikane2018PASJ,Harikane2018ApJ,Sobral2018_SC4K} now enables detailed studies of the properties of the earliest stellar populations and black holes. Studies based on the UV slopes ($\beta$) of high redshift galaxies indicate that they are consistent with little dust \citep[e.g.][]{Dunlop2012,Bouwens2014slope,Wilkins2016}. However, results regarding the nature of {the} underlying stellar populations are ambiguous due to possible contributions from nebular continuum and {dust-age-metallicity degeneracies \citep[e.g.][]{Raiter2010,Barros2014}; see also \cite{Popping2017}}. These degeneracies can only be overcome by direct spectroscopic observations that trace different states of the inter-stellar medium (ISM), but such observations have so far been limited, due to the faintness of sources.

Bright targets from wide-field ground-based surveys \citep[e.g.][]{Bowler2014,Matthee2015,Hu2016,Santos2016,Zheng2017,Jiang2017,Shibuya2017} provide unique opportunities to obtain the first detailed and resolved studies of sources within the epoch of re-ionisation. These bright sources are particularly suitable for follow-up with ALMA \citep[e.g.][]{Venemans2012,Ouchi2013,Capak2015,Maiolino2015,Smit2017,Carniani2018}. While some sources seem to be relatively dust free \citep[e.g.][]{Ota2014,Schaerer2015}, consistent with metal-poor local galaxies, others seem to already have significant amounts of dust even at $z>7$ \citep[e.g.][]{Watson2015}. Interestingly, the majority of sources is resolved in multiple components in the rest-frame UV \citep[e.g.][]{Sobral2015,Bowler2017,Matthee2017} and/or in rest-frame FIR cooling-lines \citep[e.g.][]{Maiolino2015,Carniani2017b,Matthee2017_ALMA,Jones2017_ALMA}.

In this paper we study COSMOS Redshift 7 (CR7; $z = 6.604$, L$_{\rm Ly\alpha}$=$10^{43.8}$\,erg\,s$^{-1}$; \citealt{Sobral2015}; hereafter \citetalias{Sobral2015}), a remarkably luminous source within the epoch of re-ionisation. CR7 was identified as a luminous Ly$\alpha$ candidate by \cite{Matthee2015}, while its UV counterpart was independently found as a bright, but unreliable, $z\sim6$ Lyman-break candidate \citep{Bowler2012,Bowler2014}. CR7 was spectroscopically confirmed as a luminous Ly$\alpha$ emitter by \citetalias{Sobral2015} through the presence of a narrow, high EW Ly$\alpha$ line (FWHM\,$\approx270$\,km s$^{-1}$; EW$_0\approx200$\,{\AA}). \citetalias{Sobral2015} estimated that its Ly$\alpha$ luminosity was roughly double of what had been computed in \cite{Matthee2015}, due to the Ly$\alpha$ line being detected at $\sim50$\% transmission of the narrow-band filter used in \cite{Matthee2015}.

One of the reasons that made CR7 an unreliable $z\sim6-7$ candidate Lyman-break galaxy (LBG) was the presence of an apparent $J$ band excess of roughly $\sim3$\,$\sigma$ \citep{Bowler2012,Bowler2014} based on UltraVISTA DR2 data \citepalias{Sobral2015} and the strong Ly$\alpha$ contamination in the $z$ band. The spectroscopic confirmation of CR7 as a Ly$\alpha$ emitter at $z=6.6$ and the NIR photometry provided strong hints that an emission line should be contributing to the flux in the NIR. The shallow X-SHOOTER spectra of CR7 revealed an emission line in the $J$ band (EW$_0\gsim20$\,{\AA}), interpreted as narrow He{\sc ii1640\AA} ($v_{\rm FWHM}=130$\,km\,s$^{-1}$), while no metal line was found at the current observational limits in the UV \citepalias{Sobral2015}. Such observations made CR7 unique, not only because it became the most luminous Ly$\alpha$ emitter at high redshift, but also due to being a candidate for a very low metallicity star-burst (``PopIII-like'') or AGN, particularly due to the high He{\sc ii}/Ly$\alpha\approx0.2$ line ratio estimated from photometry. As discussed in \citetalias{Sobral2015}, any `normal' metallicity source would have been detected in C{\sc iv} or C{\sc iii}] \citep[e.g.][]{Stark2015,StarkCIV,Sobral2018b}, indicating that the metallicity of CR7 should be very low \citep[e.g.][]{Hartwig2016}. As the ionisation energy of He{\sc ii} is 54.4 eV, the ionising source leading to He{\sc ii} in CR7 must be very hot, with an expected effective temperature of $T\sim10^5$K, hotter than normal stellar populations.

Due to its unique properties, CR7 has been discussed in several studies, some focusing on one of the hypotheses discussed in \citetalias{Sobral2015} that it could harbour a direct collapse black hole
\citep[DCBH, e.g.][]{Pallottini2015,Hartwig2016,Smith2016,Agarwal2016,Agarwal2017A,Pacucci2017}. However, as \cite{Dijkstra_G_DS2016} shows, the DCBH interpretation has significant problems and realistically it cannot be favoured over e.g. PopIII-like (i.e. very low metallicity; e.g. \citealt{Visbal2016,Visbal2017}) stellar populations. \cite{Dijkstra_G_DS2016} also argued that CR7's Ly$\alpha$ line is well explained by outflowing shell models, similarly to lower redshift Ly$\alpha$ emitters \citep[e.g.][]{Karman2017,Gronke2017}.

CR7 has been found to have a 3.6\,$\mu$m excess, discussed as potential e.g. H$\beta$+[O{\sc iii}]5007 emission for the source as a whole \citep{Matthee2015,Bowler2017CR7,Harikane2018ApJ}. Recent studies went beyond the direct photometric analysis presented in \citetalias{Sobral2015} and de-convolved {\it Spitzer}/IRAC data \citep[][]{Agarwal2016,Bowler2017CR7}, attempting to measure the properties of CR7's three different UV clumps. Such studies have reached similar observational results but often contradictory interpretations. For example, \cite{Bowler2017CR7} identifies the brightest UV clump in CR7 (clump A) as the brightest at 3.6\,$\mu$m and interprets such brightness as [O{\sc iii}]\,5007 emission, using it to argue for a very low metallicity population with significant binary contribution, or a low metallicity AGN. Others \citep[e.g.][]{Agarwal2017A,Pacucci2017} argue that those are the signatures of a ``post-DCBH". \cite{Bowler2017CR7} also notes {that CR7's $J$ magnitude has changed by $\approx+0.2$\,mag from the public DR2 data} used in \citetalias{Sobral2015}, which makes the SED signature for He{\sc ii} based on photometry {less significant}. \cite{Shibuya2017_spec} presented spectroscopic results of luminous Ly$\alpha$ emitters, and analysed X-SHOOTER data for CR7 to reach the same conclusions as \citetalias{Sobral2015} regarding Ly$\alpha$, but argue against the He{\sc ii} line detection. More recently, [C{\sc ii}] was detected in each of CR7's clumps with {ALMA \citep[][hereafter \citetalias{Matthee2017_ALMA}]{Matthee2017_ALMA}, with hints of a spectroscopically-backed multiple major-merger in CR7.}

In this paper, we explore new {\it HST}/WFC3 resolved grism and imaging data, re-analyse and re-interpret previous spectroscopic data to further unveil the nature of CR7. In \S\ref{Obs_and_data} we present the observations, data reduction and re-analysis of spectroscopic data. {Results are presented in \S\ref{Results}. We use the best constraints on rest-frame UV emission lines and interpret them with our {\sc cloudy} modelling in \S\ref{JWST_Predictions}. We discuss the results in} \S\ref{Discussion} and present the conclusions in \S\ref{Conclusion}. Throughout this paper, we use {AB magnitudes \citep[][]{Oke1983}}, a \cite{Salpeter1955} IMF and a $\Lambda$CDM cosmology with $H_0=70$\,km\,s$^{-1}$\,Mpc$^{-1}$, $\Omega_{\rm M}=0.3$ and $\Omega_{\Lambda}=0.7$.


\section{Observations of CR7}\label{Obs_and_data}

%
%
\begin{figure}
\includegraphics[width=8.6cm]{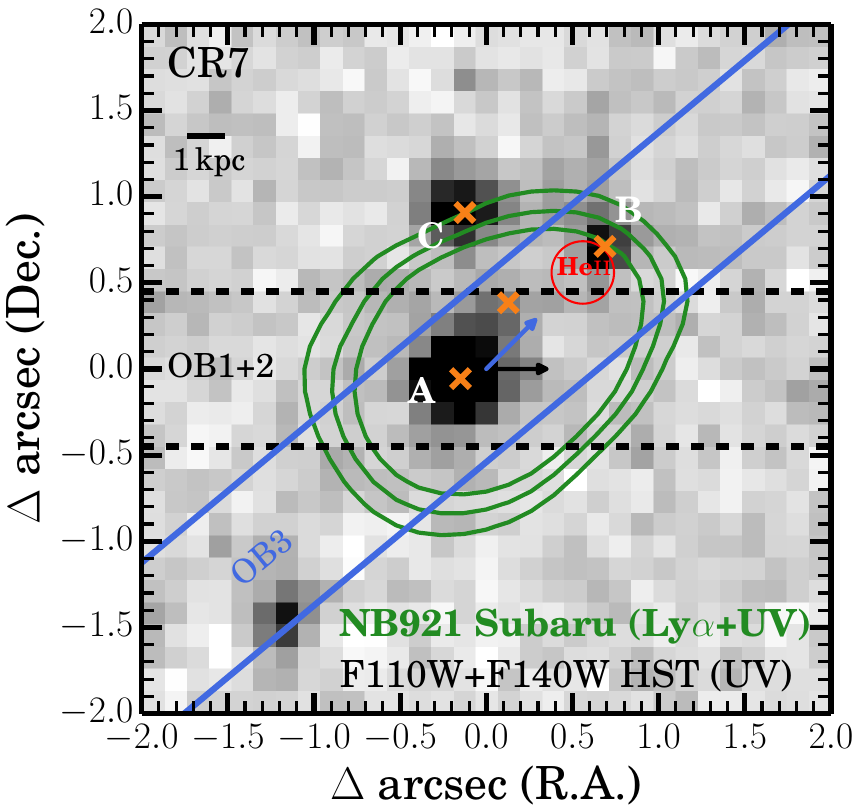}
\caption{The {\it HST}/WFC3 stacked image showing the rest-frame UV {(contrast cutoffs: $-1\sigma$ and 5$\sigma$)}, and the NB921 ground-based Ly$\alpha$ contours {(3, 4, 5\,$\sigma$)} of CR7 \citep[][]{Matthee2015,Sobral2015}. We also show the approximate position, rotation and on-sky width (0.9$''$) of the {X-SHOOTER slit used for the 3 OBs (see \S\ref{xshoot_obs}). The two arrows point towards positive spatial locations in the reduced 2D spectra, i.e., positive offsets in the Y coordinate of the reduced {2D spectra (see e.g. Figure \ref{Xshooter_2D_Lya_per_OB})}. The location of He{\sc ii} detected in OB3 is also indicated based on the $+0.8''$ offset from the central position, making it consistent with being towards clump B but not on top of the UV clump. The orange crosses indicate the positions we use to place apertures on individual clumps or for the full system.}}
\label{XSHOOTER_angle}
\end{figure}

\subsection{Imaging Observations and SFR properties from {\it HST} and ALMA}\label{imaging_obs}

{\it HST} imaging reveals that CR7 consists of three ``clumps" \citep[][]{Sobral2015,Bowler2017}; see Figure \ref{XSHOOTER_angle}. We note that slit spectroscopic follow-up was targeted roughly at the peak of Ly$\alpha$ flux, and thus roughly at the position of clump A (see Figure \ref{XSHOOTER_angle}), but without knowing that the source could be resolved in 3 UV clumps \citepalias[see][]{Sobral2015}. Therefore, clumps B and C were not originally spectroscopically confirmed even though they are within the Ly$\alpha$ halo as observed with the narrow-band data and have a Lyman-break consistent with $z>6$. Deep, high spatial and spectral resolution ALMA [C{\sc ii}] data have nonetheless allowed to spectroscopically confirm each of the UV clumps A, B and C as being part of the same system \citepalias{Matthee2017_ALMA}. {Readers are referred to \citetalias{Matthee2017_ALMA} for a discussion on the spectroscopic confirmation of both clumps B and C and on the further dynamical and physical information inferred from the ALMA data, including discussions on the extra [C{\sc ii}] component between clumps B and C (M$_{\rm dyn}\sim2\times10^{10}$\,M$_{\odot}$; C-2 in \citetalias{Matthee2017_ALMA}) which is not seen in the UV \citep[see also][]{Carniani2017b}.}

Clump A, the brightest (M$_{UV} = -21.6\pm0.1$; \citetalias{Matthee2017_ALMA}), roughly coincides with the peak of Ly$\alpha$ emission and has a UV slope $\beta$ (corrected for the contribution of Ly$\alpha$ to the F110W photometry) of $\beta=-2.3\pm0.4$ (measured within a 1$''$ diameter aperture; \citetalias{Matthee2017_ALMA}). Clumps B and C are fainter (M$_{UV} = -19.8\pm0.2$ and M$_{UV} = -20.1\pm0.1$, respectively; Figure \ref{XSHOOTER_angle}) and show $\beta = -1.0\pm1.0$ and $-2.3\pm0.8$ in 0.4$''$ apertures \citep[see also][]{Bowler2017CR7}. As the UV slopes are quite uncertain, they allow for large dust attenuations and hence uncertain SFRs. However, as shown in \citetalias{Matthee2017_ALMA}, constraints on the IR continuum luminosity from very deep ALMA observations of CR7 can mitigate these uncertainties. In practice, as CR7 is undetected in dust continuum, it implies a relatively low FIR luminosity of L$_{\rm IR}$(T$_{d}=35$\,K)\,$<3.1\times10^{10}$\,L$_{\odot}$ and a dust mass M$_{\rm dust}<8.1\times10^{6}$\,M$_{\odot}$ (3\,$\sigma$ limits). Such limits imply a maximum dust obscured star formation rate of $<5.4$\,M$_{\odot}$\,yr$^{-1}$ for the full system. Overall, the combination of {\it HST} and ALMA observations reveal dust-corrected SFR$_{\rm UV+IR} = 28^{+2}_{-1},  5^{+2}_{-1},  7^{+1}_{-1}$\,M$_{\odot}$ yr$^{-1}$ \citepalias[see][]{Matthee2017_ALMA} for clumps A, B and C, respectively, for a Salpeter IMF (and a factor $\approx1.8$ lower for a Chabrier IMF). The SFR of the full CR7 system (A,B,C) is $45^{+2}_{-2}$\,M$_{\odot}$ yr$^{-1}$, {taking into account the ALMA constraints for obscured SFR.}

\subsection{Re-analysis of X-SHOOTER observations}\label{xshoot_obs}

We re-analyse the X-SHOOTER data originally presented in \citetalias{Sobral2015}. {The NIR spectroscopic data in \citetalias{Sobral2015} were flux-calibrated using public DR2 UltraVISTA $J$ band photometry. Those public data revealed a strong $J$ band excess for CR7 \citepalias{Sobral2015}. More recently, \cite{Bowler2017CR7} used DR3 data {to measure a fainter J band magnitude}, due to a change from DR2 to DR3 in the public UltraVISTA $J$ band photometry. We investigate such potential change in UltraVISTA $J$ band data separately in Section \ref{Ultravista_variability}.}

The VLT/X-SHOOTER data were obtained over 3 different observing blocks (OBs; see Figure \ref{XSHOOTER_angle}) of about 1 hour each, with two OBs obtained on 22 January 2015 {(seeing $1.2''$; varying from 0.8$''$ to 1.6$''$)} and a final OB (a repeat of OB1, which we name OB3 in this paper, but that is formally called `OB1' in the ESO archive). OB3 was obtained with {a seeing of $0.8''$, varying from 0.7$''$ to 0.9$''$, and thus in better conditions than OBs 1 and 2 and was} done on 15 February 2015. We reduce all OBs separately. All OBs used a 0.9$''$ slit in both the VIS and NIR arms.

For the first two OBs a PA angle of 0\,deg was used (see Figure \ref{XSHOOTER_angle}), together with an acquisition source at 10:01:03.156 $+$01:48:47.89. Offsets of $-77.27''$ (R.A.) and $-32.63''$ (Dec.) were used to offset from the acquisition source to CR7. The acquisition for the first OB (OB1, 22 January 2015) was suspected to be relatively off-target due to an unreliable acquisition star centring (acquisition star was not centred in the slit), leading to an apparent lower Ly$\alpha$ flux and a spatially truncated and complex/double peaked Ly$\alpha$ profile, different from that found in the OB2 which was done with a good acquisition and with {{Keck/DEIMOS} data (see Figure \ref{Xshooter_2D_Lya_per_OB} and \citetalias{Sobral2015}). When repeating OB1 and in order to avoid problems with acquisition, another acquisition source was used: 10:01:00.227, 01:48:42.99, applying an offset of $-33.34''$ (R.A.) and $-27.74''$ (Dec.) and this time with a PA angle of $-39.76$\,deg, {in order to align the slit with the elongation of the Ly$\alpha$ 2D distribution obtained from the narrow-band imaging\footnote{At the time of preparation of all spectroscopic observations of CR7 in 2014 and early 2015 (and the multi-wavelength analysis) the resolved nature of CR7, only revealed by {\it HST} data in April 2015, was unknown.} (Figure \ref{XSHOOTER_angle})}.

We use the X-SHOOTER pipeline (v2.4.8; \citealt{Modigliani2010}), and follow the steps fully described in {\cite{Matthee2017} and \cite{Sobral2018b}}, including flux calibration. We note that our data reduction results in a significantly improved wavelength calibration in the NIR arm when compared to \citetalias{Sobral2015}, {which we find to be off by $-6.9\pm0.6$\,{\AA} ($\lambda_{\rm air}$) in the NIR arm when compared to our reduction\footnote{It is important to note that in the literature $\lambda_{\rm air}$ can be used instead of $\lambda_{\rm vacuum}$ and that He{\sc ii} is sometimes used as 1640.0\,{\AA} instead of 1640.47\,{\AA} in vacuum; these can combine to lead to multiple offsets between different studies. Such small differences are typically negligible at lower redshift and for low resolution spectra, but they become important at high redshift and for high resolution spectra, as they can lead to significant discrepancies and offsets.}; this is obtained by matching OH lines (see Figure \ref{OH_offset_lines})}. We find this offset to be due to the use of old arcs in \citetalias{Sobral2015}. The latest ESO public reduction and \cite{Shibuya2017_spec} obtain the same wavelength calibration as us using the most up-to-date pipeline. In the VIS arm we find no significant differences in the wavelength calibration when comparing to \citetalias{Sobral2015}, but we now flux calibrate the data (using appropriate telluric stars) without relying on any narrow- or broad-band photometry, unlike \citetalias{Sobral2015}. In Figure \ref{Xshooter_2D_Lya_per_OB} we show the reduced 2D spectra centred on Ly$\alpha$ for each individual OB (note that the positive spatial direction is indicated with an arrow in Figure \ref{XSHOOTER_angle}). {We also show the combined stack of the 3 OBs and when combining only the 2 first OBs which trace a different spatial region when compared to OB3}. We present the results in Section \ref{XSHOOT_results}. 

{Our reduced spectra show a spectral resolution (FWHM based on sky lines) of $\approx1.6$\,{\AA} at $\approx9000$\,{\AA} ($\approx55$\,km\,s$^{-1}$), corresponding to $R\sim5600$ and $\approx3.5$\,{\AA} at $\approx$\,16,000\,{\AA} ($\approx65$\,km\,s$^{-1}$), corresponding to $R\sim4600$. In order to improve the signal-to-noise and reduce noise spikes and prevent the dominance of individual pixels, we bin our 1D spectra to 1/3 of the resolution by using bins of 0.6\,{\AA} in the VIS and 1.2\,{\AA} in the NIR arm. We use these 1D spectra converted to $\lambda_{\rm vacuum}$ throughout our analysis unless noted otherwise. The analysis is done following \cite{Sobral2018b} using Monte Carlo (MC) forward modelling to search for emission lines and measure the uncertainties. We provide further details in relevant sections throughout the manuscript.}

\begin{figure}
\includegraphics[width=8.6cm]{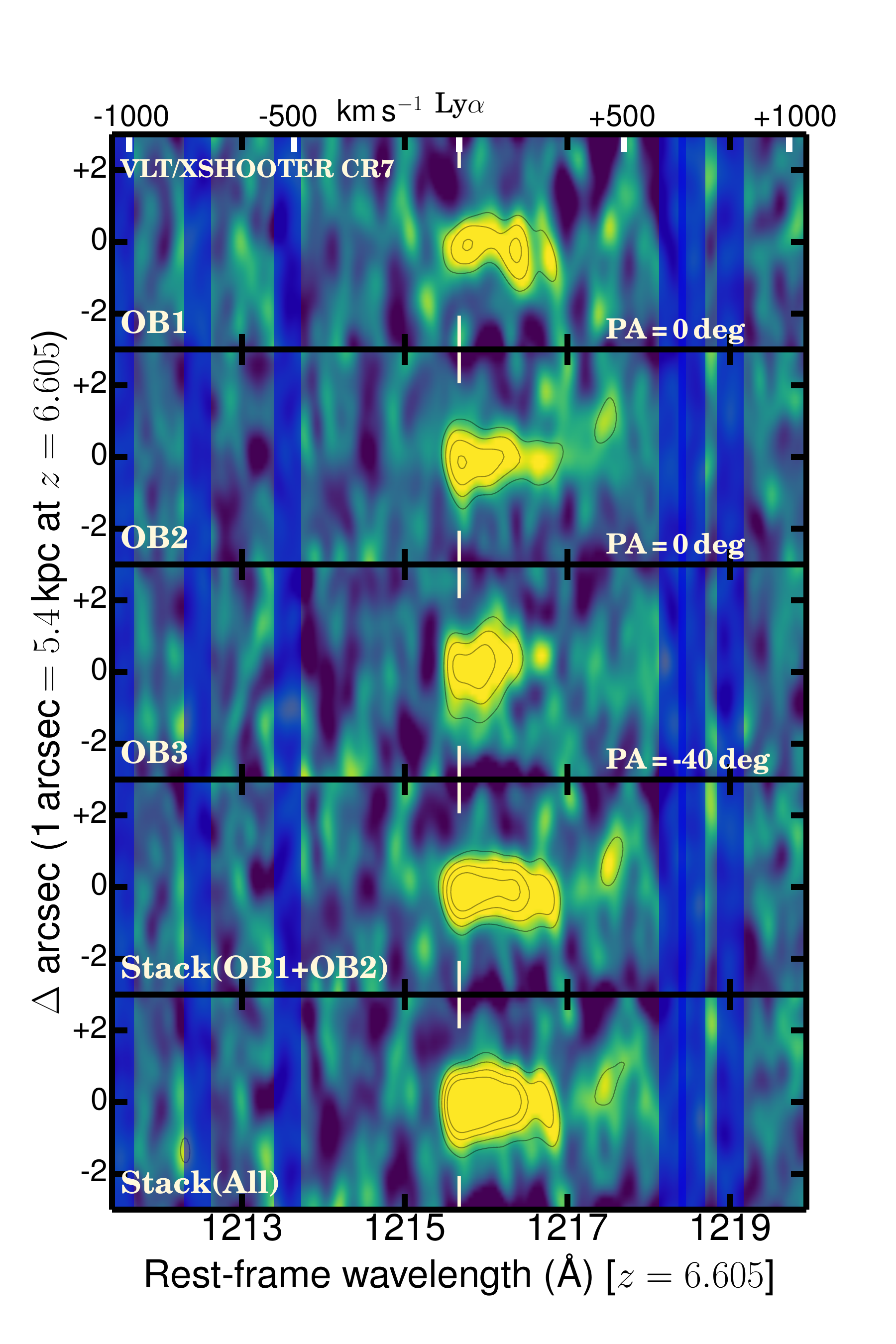}
\caption{Our reduced and flux calibrated 2D X-SHOOTER spectra, zoomed-in at Ly$\alpha$, in {S/N space showing 2, 3, 4 and 5\,$\sigma$ contours after smoothing with a 3 spectral-spatial pixel Gaussian kernel. The location of sky lines are shown, even though all these are relatively weak. OB1 and OB2 were done consecutively on the same night but OB2 resulted from a better acquisition of the offset star; both were done under variable seeing. OB3 was done with a different slit angle, sampling along the axis of clumps A and B (see Figure \ref{XSHOOTER_angle}) and under better and more stable seeing conditions.}}
\label{Xshooter_2D_Lya_per_OB}
\end{figure}

\subsection{Re-analysis of SINFONI observations}\label{sinfoni_obs}

{We also re-reduce the SINFONI data presented in \citetalias{Sobral2015}. The final data-cube in \citetalias{Sobral2015} was produced with equal weights for all exposures by using the SINFONI pipeline to reduce all the OBs together with a single set of calibration observations.} The data were {scaled using the $J$ magnitude from UltraVISTA} and the flux implied for He{\sc ii} from UltraVISTA. Finally, the stack was combined with X-SHOOTER data which had a systematic offset in wavelength {of $6.9$\,{\AA}}, as stated in Section \ref{xshoot_obs}.

CR7 was observed with SINFONI in Mar-Apr 2015 (program 294.A-5039) with 6 different OBs of about 1 hour each. Four of those OBs were classed A (highest quality), one of them was classed B (seeing $>1''$) and another one was classed C (bad quality, due to clouds). Here we neglect the one classed C.

We use the SINFONI pipeline v.2.5.2 and implement all the steps using {\sc esorex}. We reduce each OB with the appropriate specific calibration files, done either on the same night or on the closest night possible. We reduce each OB individually, along with each standard/telluric star. In total, 5 different telluric stars were observed, 1 per OB/night of observations, and we reduce those observations in the same way as the science observations. In order to flux calibrate we use 2MASS $JHK$ magnitudes of each star. We extract the standard stars' spectra by obtaining the total counts per wavelength (normalised by exposure time) in the full detector, following the procedure in the pipeline, and we then re-extract them over the apertures used to extract the science spectra. This allows us to derive aperture corrections which vary per OB (due to seeing), which are typically $\sim$1.5 for 1.4$''$ extraction apertures, and $\sim$1.2 for 2$''$ aperture extractions.

We find that the absolute astrometry of the pipeline reduced data-cubes is not reliable, as each OB (which is done with the same offset star and with the same jitter pattern) results in shifts of several arcsec between each reduced data-cube. We attempt to extract spectra in the R.A. and Dec. positions of CR7 assuming the astrometry is correct but fail to detect any signal, with the stacked spectra resulting in {high noise levels due to the extraction away from the centre}. Finally, we make the assumption that the data cubes are centred at the position of the first exposure which serves as reference for the stack of each OB, and extract 1D spectra per OB with apertures of 0.9$''$, 1.4$''$ and 2$''$ (using our aperture corrections), which we assume are centred at the peak of Ly$\alpha$ emission and will be able to cover the full CR7 system. {In order to improve our sky subtraction, we compute the median of 1,000 empty apertures with the same size as the extraction aperture and subtract it from the extraction aperture. We also use the 1,000 apertures per spectral element to compute the standard deviation and use it as the noise at that specific wavelength}. Finally, we stack spectra from the different OBs by weighting them with the inverse of the {variance ($\sigma^2$)}. Reduced SINFONI spectra have a resolution (FWHM, based on OH lines) of $\sim6.4$\,{\AA} at $\sim1.2$\,$\mu$m ($R\sim1900$; $\sim150$\,km\,s$^{-1}$). When binned to {1/3} of the resolution, the spectra ($0.9''$ apertures, stacked) reach a 1$\sigma$ {flux limit of $\approx5\times10^{-19}$\,erg\,s$^{-1}$\,cm$^{-2}$\,{\AA}$^{-1}$ away from OH sky lines at an observed $\lambda\approx1.245$\,$\mu$m.}

%
%
\begin{figure}
\includegraphics[width=8.4cm]{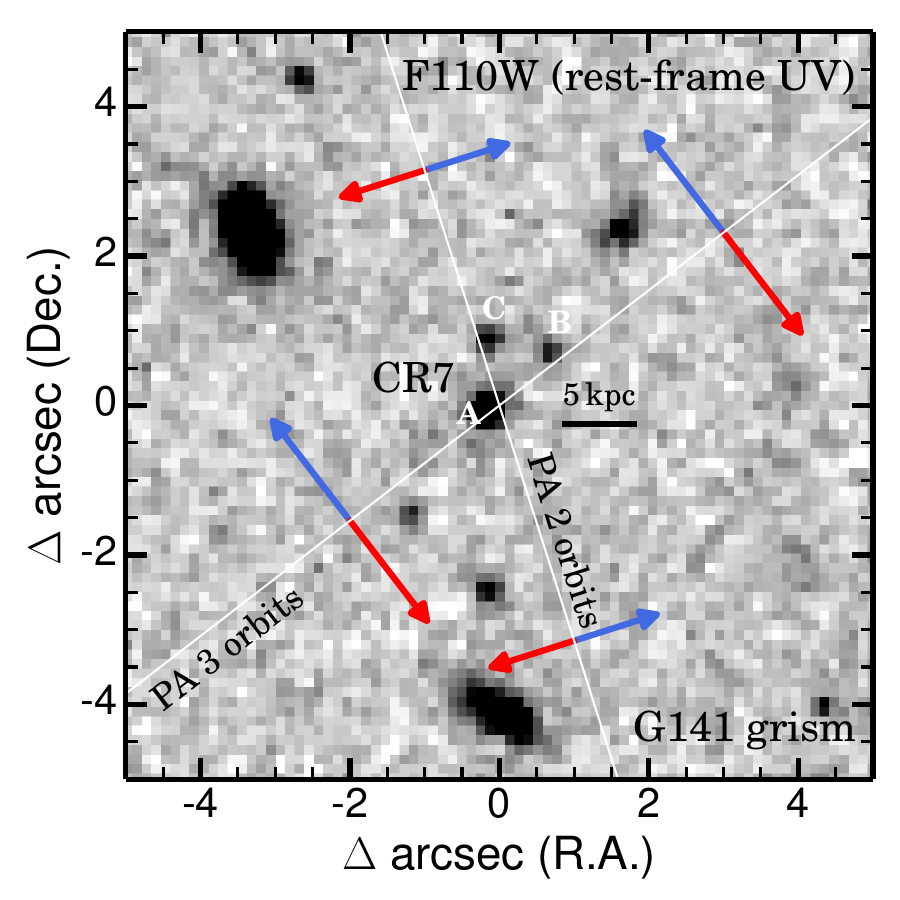}
\caption{HST/WFC3 F110W ($Y+J$) image centred on CR7 and the immediate surroundings for our G141 grism observations. We indicate the PA angles used for each of the 2 visits done: one observing for 2 orbits and the final one observing for 3 orbits. We also indicate the dispersion direction and the direction in which bluer/redder light gets dispersed once the grism is used to take observations. Our observations allow us to avoid contamination from nearby sources and obtain spectra for each of the components A, B and C for CR7. {We also show the 5\,kpc scale at $z=6.6$.}}
\label{Grism_angles}
\end{figure}

\subsection{WFC3/{\it HST} grism Observations}\label{grism_obs}

We observed CR7 with the WFC3 grism with GO program 14495 (PI: Sobral). Observations were conducted over a total of 5 orbits: 2 orbits during 21 Jan 2017 and 3 further orbits conducted during 17 Mar 2017. We used two different PA angles (252.37\,deg and 322.37\,deg; see Figure \ref{Grism_angles}), each calculated to avoid significant contamination by nearby bright sources and in order to investigate the spectra of {the rest-frame UV components} A, B and C separately.

For each orbit, we obtained an image with the F140W filter, two grism observations (dithered) with the G141 grating (central wavelength 13886.72\,{\AA}), and another image after the second grism observation. These allow us to correctly identify the sources and to clearly locate {the rest-frame UV} clumps A, B and C within CR7. The F140W images were obtained at the start and end of each orbit with the aim to minimize the impact of variable sky background on the grism exposure (due to the bright Earth limb and the He\,1.083\,$\mu$m line emission from the upper atmosphere; see \citealt{Brammer2014}). A four-point dithering pattern was used to improve the sampling of the point-spread function and to overcome cosmetic defects of the detector.

We obtained imaging exposures of 0.25\,ks and grism exposures of 1.10\,ks. Our total exposure grism time with G141 is 11.0\,ks. For a full description of the calibration of the WFC3/G141 grism, see e.g. \cite{Kuntschner2010}.

\subsubsection{Data reduction and extraction}\label{grism_obs}

We reduce the data following \cite{Brammer2012}. The grism data were reduced using the grism reduction pipeline developed by the 3D-{\it HST} team \citep[e.g.][]{Brammer2012,Momcheva2016}. The main reduction steps are fully explained in \cite{Momcheva2016}. In summary, the flat-fielded and global background-subtracted grism images are interlaced to produce 2D spectra for each of the UV clumps A, B and C, independently. We also identify any potential contamination from faint and/or nearby sources and subtract it when we extract the 1D spectra. {Our reduced data show a resolution of $R\sim100$ (FHWM 150\,{\AA}) at $\lambda\sim1.2$\,$\mu$m ($\approx3750$\,km\,s$^{-1}$), and thus a resolution of $\sim20$\,{\AA} at $\sim1600$\,{\AA} rest-frame for CR7 ($z=6.6$). We bin the data to 1/3 of the resolution ($\approx50$\,\AA, observed). We note that the {\it HST}/WFC3 grism resolution is $\approx40$ times worse than X-SHOOTER at $\lambda\sim1.2$\,$\mu$m.}

We extract the spectra of the 3 major components of CR7 from their central positions by using the rest-frame UV continuum images obtained with {\it HST}. We see clear continuum in the 2D spectrum for clump A (the brightest) and weak continuum from B. We find that apart from some minor contamination at observed $\lambda\sim15500-15700$\,{\AA}, the spectra of the 3 clumps of CR7 are not contaminated by any other nearby sources, as expected from our observing planning (Figure \ref{Grism_angles}). We thus estimate the noise on the CR7 spectrum by extracting spectra in a range of spatial locations (per clump) with similarly low contamination. We use the standard deviation per wavelength as the estimate of our 1\,$\sigma$ error and we use these to quantify the signal to noise and to evaluate the significance of both the continuum and the detection of any emission lines. {Our 1D spectra for the extraction of the 3 components of CR7 show an average noise level of $(3.1-3.4)\times10^{-19}$\,erg\,s$^{-1}$\,cm$^{-2}$\,{\AA}$^{-1}$ for $1.1<\lambda_{\rm observed}<1.6$\,$\mu$m.}

\section{Results}\label{Results}

\begin{figure}
\includegraphics[width=8.6cm]{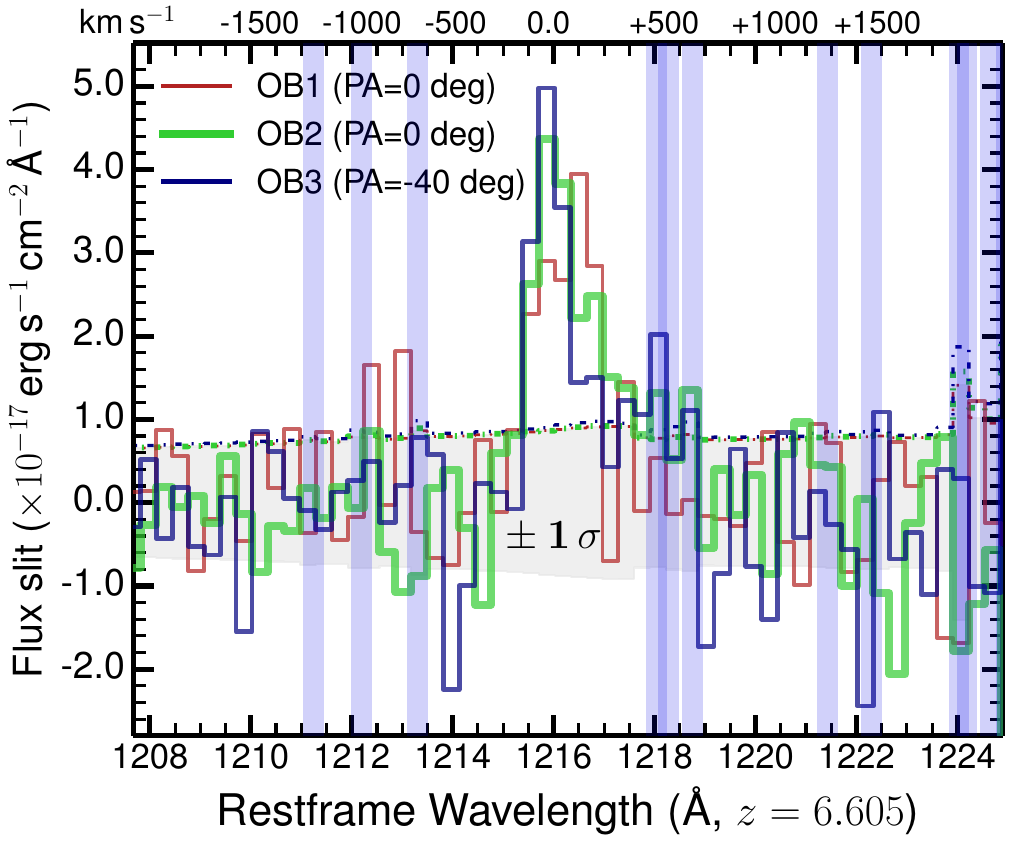}
\caption{The extracted 1D spectra from X-SHOOTER at the position of Ly$\alpha$ showing results from different OBs which trace different spatial scales and different angles for CR7 (see Figure \ref{XSHOOTER_angle}). We show spectra binned by 75\,km\,s$^{-1}$. We find that OB3, that traces along the Ly$\alpha$ major axis, connecting A to B, shows the highest flux peak and the narrowest Ly$\alpha$ profile, {with a FWHM of $180^{+40}_{-30}$\,km\,s$^{-1}$. Both OB1 and OB2, obtained with a 0\,deg PA angle show a broader Ly$\alpha$ profile than OB3. The differences between OB1 or OB2 and OB3 are only significant at the $1.7-1.8$\,$\sigma$ level individually, but the stack of OB1 and OB2 yields a Ly$\alpha$ FWHM which is $\approx3$\,$\sigma$ away from that of OB3 (see Table \ref{XSHOOT_MC}).}}
\label{XSHOOTER_1D_Lya}
\end{figure} 

\subsection{ VLT spectroscopy}\label{XSHOOT_results}

\subsubsection{ Ly$\alpha$ in X-SHOOTER}

In Figure \ref{Xshooter_2D_Lya_per_OB} we show the 2D spectra for our re-analysis of the X-SHOOTER data, in a signal-to-noise scale, focusing on Ly$\alpha$. We find potential variations in the Ly$\alpha$ profile, indicating that we {may be} probing different spatial regions within the source. This is likely due to the bad acquisition for OB1 (in comparison to OB2; {both OBs were done with variable seeing of $\sim$1.2$''$}) and due to a different acquisition star and PA angle for OB3. {Even though the S/N is not high enough for a robust conclusion}, OB3 suggests a redshifted component of Ly$\alpha$ in the direction of clump B (see Figure \ref{XSHOOTER_angle}). As can be seen in more detail in Figure \ref{XSHOOTER_1D_Lya}, OB3 reveals a narrower Ly$\alpha$ profile ($\sim180$\,km\,s$^{-1}$) than OB2 ($\sim310$\,km\,s$^{-1}$), hinting that the Ly$\alpha$ FWHM may be narrower along the major axis of Ly$\alpha$ (running from A to B), but both OB2 and OB3 show the same/similar blue cut-off. In order to quantify any differences in the Ly$\alpha$ profile, we perform a Monte Carlo simulation, perturbing each spectral element in the 1D spectra (1/3 of the resolution) within its Gaussian distribution uncertainty independently. We do this 10,000 times \citep[following the methodology in][]{Sobral2018b} and each time we measure the FWHM of the Ly$\alpha$ line by fitting a Gaussian and deconvolve it with the resolution. Results are given in Table \ref{XSHOOT_MC}. We find that OB1 and OB2 yield Ly$\alpha$ FWHMs of $290^{+62}_{-45}$\,km\,s$^{-1}$ and $310^{+95}_{-67}$\,km\,s$^{-1}$, respectively, while for OB3 we obtain a narrower Ly$\alpha$ profile of $177^{+44}_{-30}$\,km\,s$^{-1}$ and for the stack of all OBs we obtain $270^{+35}_{-30}$\,km\,s$^{-1}$, in agreeement with \citetalias{Sobral2015}. Our results suggest that there may be a difference between the profile of Ly$\alpha$ between a PA angle of 0 (tracing just clump A) and a PA angle of 40 that connects clumps A and B. Such differences between OB1 or OB2 and OB3 are only significant at the $1.7-1.8$\,$\sigma$ level individually, but the difference between OB3 and the stack of OB1 and OB2 is at the $\approx3$\,$\sigma$ level. Deeper data are needed to fully confirm these potential spatial differences in the Ly$\alpha$ profile.

Interestingly, \citetalias{Matthee2017_ALMA} finds that the axis perpendicular to the Ly$\alpha$ major axis shows the largest velocity shift in [C{\sc ii}], from the most blueshift towards C to the highest redshift towards the opposite direction, and with a total velocity shift of $\sim300$\,km\,s$^{-1}$, similar to the Ly$\alpha$ FWHM in OB2 (Figure \ref{XSHOOTER_1D_Lya}). It may well be that Ly$\alpha$ itself is tracing complex dynamics, or that we are seeing more complex radiation transfer effects or different H{\sc i} column densities. {Deep observations with MUSE on the VLT and further modelling \citep[e.g.][]{Gronke2017,Matthee2018_COLA} will robustly clarify the current open scenarios.}

\subsubsection{{HeII in X-SHOOTER}}\label{XSHOT_HeII}

\begin{figure}
\includegraphics[width=8.6cm]{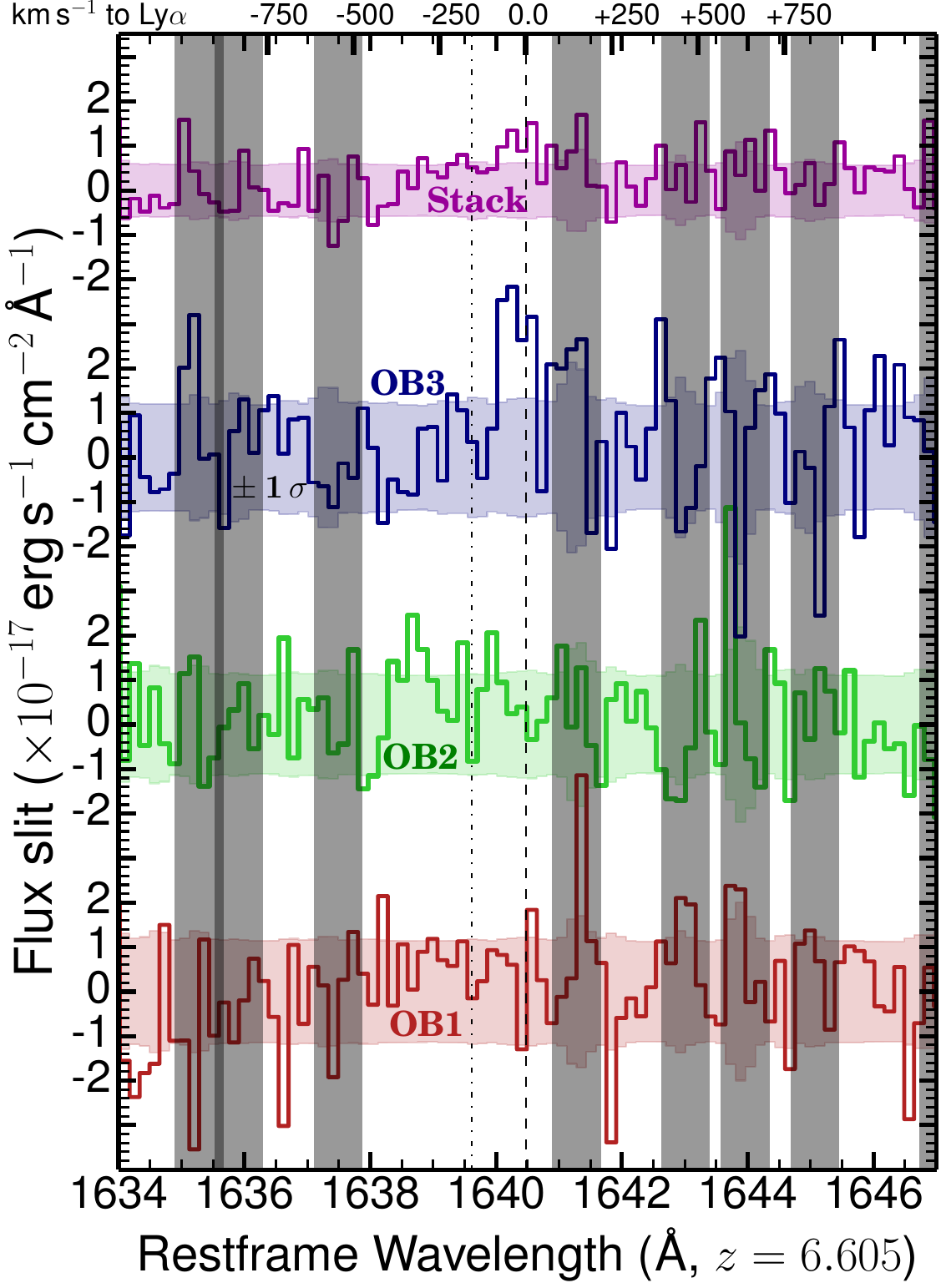}
\caption{{The extracted 1D spectra from our X-SHOOTER re-analysis of individual OBs and the full stack at the expected location of He{\sc ii}. OH lines are clearly labelled. We find no significant He{\sc ii} detection for CR7 in the spatial locations covered by OB1 and OB2. OB3 reveals a significant He{\sc ii} detection (which dominates the signal in \citetalias{Sobral2015}), explaining the detection in the full stack. We show the expected location of the He{\sc ii} line in the case of no velocity shift from Ly$\alpha$ and also where we would expect to detect based on {[C{\sc ii}]}-ALMA emission from clump A (dot-dashed). We find that the He{\sc ii} signal is consistent with a relatively small velocity offset from Ly$\alpha$ of $\sim100$\,km\,s$^{-1}$, although we note that the line is spatially coincident in OB3 with a redshifted Ly$\alpha$ component.}}
\label{XSHOOTER_1D_HeII}
\end{figure}

We show our re-analysis of X-SHOOTER data, split by OB, in Figure \ref{XSHOOTER_1D_HeII}, where we present the extracted 1D spectra at the expected rest-frame wavelength of He{\sc ii} at $z=6.605$. The results of our MC analysis for OB3 and a comparison to \citetalias{Sobral2015} are shown in Figure \ref{XSHOOTER_1D_MCMC_HeII}. The full results for all OBs and stacks are presented in Table \ref{XSHOOT_MC}. We also present the 2D spectrum per OB in Figure \ref{Xshooter_2D_per_OB}.


Our re-analysis is able to recover the He{\sc ii} emission line detected in \citetalias{Sobral2015}, but we can show that the signal is coming from OB3\footnote{OB3 was observed with the best, most stable seeing and with the slit aligned with the major axis of the Ly$\alpha$ extent. OB3 also shows the highest Ly$\alpha$ flux peak (Figure \ref{XSHOOTER_1D_Lya}) and the narrowest Ly$\alpha$ profile.} (see Figures \ref{XSHOOTER_1D_HeII}, \ref{XSHOOTER_1D_MCMC_HeII} and \ref{Xshooter_2D_per_OB}). Based on OB3 only, we detect He{\sc ii} at a $\approx3.8$\,$\sigma$ level with a flux of $3.4^{+1.0}_{-0.9}\times10^{-17}$\,erg\,s$^{-1}$\,cm$^{-2}$ (see Table \ref{XSHOOT_MC})\footnote{Simply placing an aperture in the 2D spectra of OB3 without any binning or smoothing leads to a flux of $\approx3\times10^{-17}$\,erg\,s$^{-1}$\,cm$^{-2}$.}. The 2D spectra of OB3 also shows negatives up and down from the offsets along the slit\footnote{Splitting OB3 in different sets of exposures leads to very low S/N, but we do not find any single exposure that is dominating the signal. This means the signal is not a cosmic ray or an artefact. Nevertheless, given the low signal-to-noise from just one OB there is still the chance that some significant OH variability during the observations could have at least contributed to boosting the signal, although the errors take OH lines into account.} (Figure \ref{Xshooter_2D_per_OB}). These are typically taken as clear indications that an emission line is real. The detected He{\sc ii} line in OB3 has a measured FWHM of $210^{+70}_{-80}$\,km\,s$^{-1}$, consistent with measurements from \citetalias{Sobral2015} (see Figure \ref{XSHOOTER_1D_MCMC_HeII}). The He{\sc ii} FWHM is statistically consistent within 1\,$\sigma$ with the Ly$\alpha$ FWHM in OB3 (see Table \ref{XSHOOT_MC}). The He{\sc ii} signal from OB3 is consistent with a redshift of $z=6.604\pm0.002$, and thus implies a relatively small velocity offset from Ly$\alpha$ of  $\sim100$\,km\,s$^{-1}$ or less, being closer in velocity to the systemic redshift of clumps A or B ($z=6.601\pm0.001$; see Figure \ref{XSHOOTER_1D_HeII}), than to the slightly lower redshifts measured for the other components in the CR7 system ($z=6.593-6.600$; \citetalias{Matthee2017_ALMA}). However, while the line is spatially offset from A and is closest to the UV clump B (see Figure \ref{XSHOOTER_angle} for spatial context) it is not found to be co-located with B and thus may trace another component in the system. New observations are required to improve the flux constraints on He{\sc ii} and to locate it spatially.

\begin{figure}
\includegraphics[width=8.6cm]{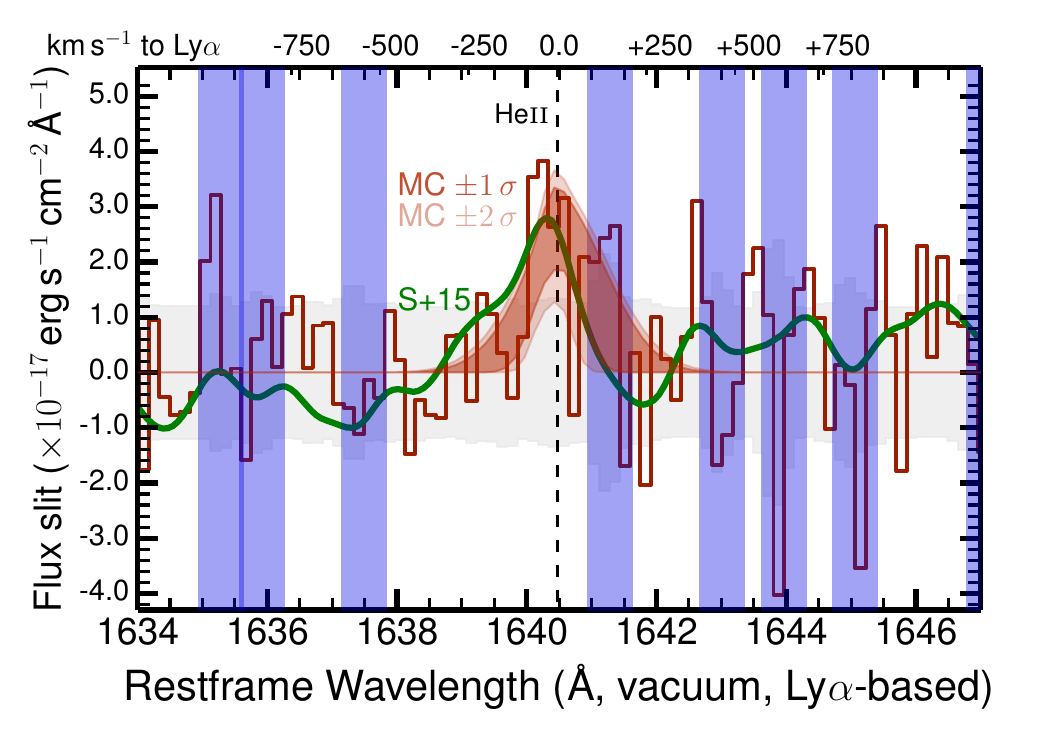}
\caption{{The spectrum of CR7 for OB3, along a PA angle of $-40$\,deg and extracted centred on the signal in the NIR, 0.8$''$ away from the peak brightness of Ly$\alpha$ towards clump B. We show the results of our forward modelling MC analysis, by perturbing the spectrum 10,000 times and the range of fits encompassing $\pm1$ and $\pm2$\,$\sigma$. We also show the location of OH/sky lines. As a comparison, we show the 1D spectra presented in \citetalias{Sobral2015}, shifted in wavelength by $+6.9$\,{\AA} and converted to vacuum and arbitrarily normalised in flux for comparison. The signal in \citetalias{Sobral2015} is consistent with being dominated by OB3, but it is smoothed with a wide Gaussian kernel and also by masking sky lines before smoothing.}}
\label{XSHOOTER_1D_MCMC_HeII}
\end{figure}

%
%
\begin{table*}
 \centering
\caption{Results of our MC measurements of X-SHOOTER CR7 spectra (following \citealt{Sobral2018b}). The results present the median values of fluxes (median of the integrated Gaussian fluxes) and the 16th and 84th percentiles as the lower and upper errors. We also present similar values for the full width at half maximum, deconvolved for resolution (FWHM) from all Gaussian fits per line. For OB1, OB2 and the stack of those OBs, He{\sc ii} is not detected above 2.5\,$\sigma$ and we provide the derived 99.4 percentile ($<2.5$\,$\sigma$) as an upper limit, but also provide the median fluxes and 16th and 84th percentiles (in brackets) for comparison. No slit corrections are applied for these specific measurements but note that such corrections are particularly important for the Ly$\alpha$ line which is spatially extended beyond what the slit captures.}
 \label{XSHOOT_MC}
\begin{tabular}{cccccc}
 \hline 
 Spectra  & PA angle  & F$_{\rm Ly\alpha}$/$10^{-17}$ & FWHM$_{\rm Ly\alpha}$ & F$_{\rm HeII}$/$10^{-17}$ & FWHM$_{\rm HeII}$  \\
 OBs/Stack  & (degree)  &  (erg\,s$^{-1}$\,cm$^{-2}$) & (km\,s$^{-1}$) & (erg\,s$^{-1}$\,cm$^{-2}$) & (km\,s$^{-1}$)  \\
 \hline
 OB1  & 0  & $4.8^{+0.7}_{-0.7}$ & $290^{+62}_{-45}$   &  $<7.8$ ($1.8^{+2.5}_{-2.0}$) & ---   \\
 OB2  & 0  & $5.9^{+1.0}_{-1.0}$ & $310^{+95}_{-67}$   &  $<5.3$ ($0.8^{+1.0}_{-0.8}$) & ---   \\
 OB3  & $-40$  & $4.4^{+0.8}_{-0.6}$ & $177^{+44}_{-30}$   &  $3.4^{+1.0}_{-0.9}$ & $210^{+70}_{-83}$   \\
 \hline
 Stack (OB1+OB2)  & 0  & $5.8^{+0.7}_{-0.6}$ & $350^{+56}_{-40}$   &  $<4.1$ ($0.8^{+0.9}_{-0.8}$) & ---   \\
 Stack (all)  & $0-40$  & $5.2^{+0.5}_{-0.4}$ & $270^{+35}_{-30}$   &  $2.0^{+0.6}_{-0.6}$ & $330^{+113}_{-120}$   \\
\hline
\end{tabular}
\end{table*}

\begin{figure*}
\includegraphics[width=16.7cm]{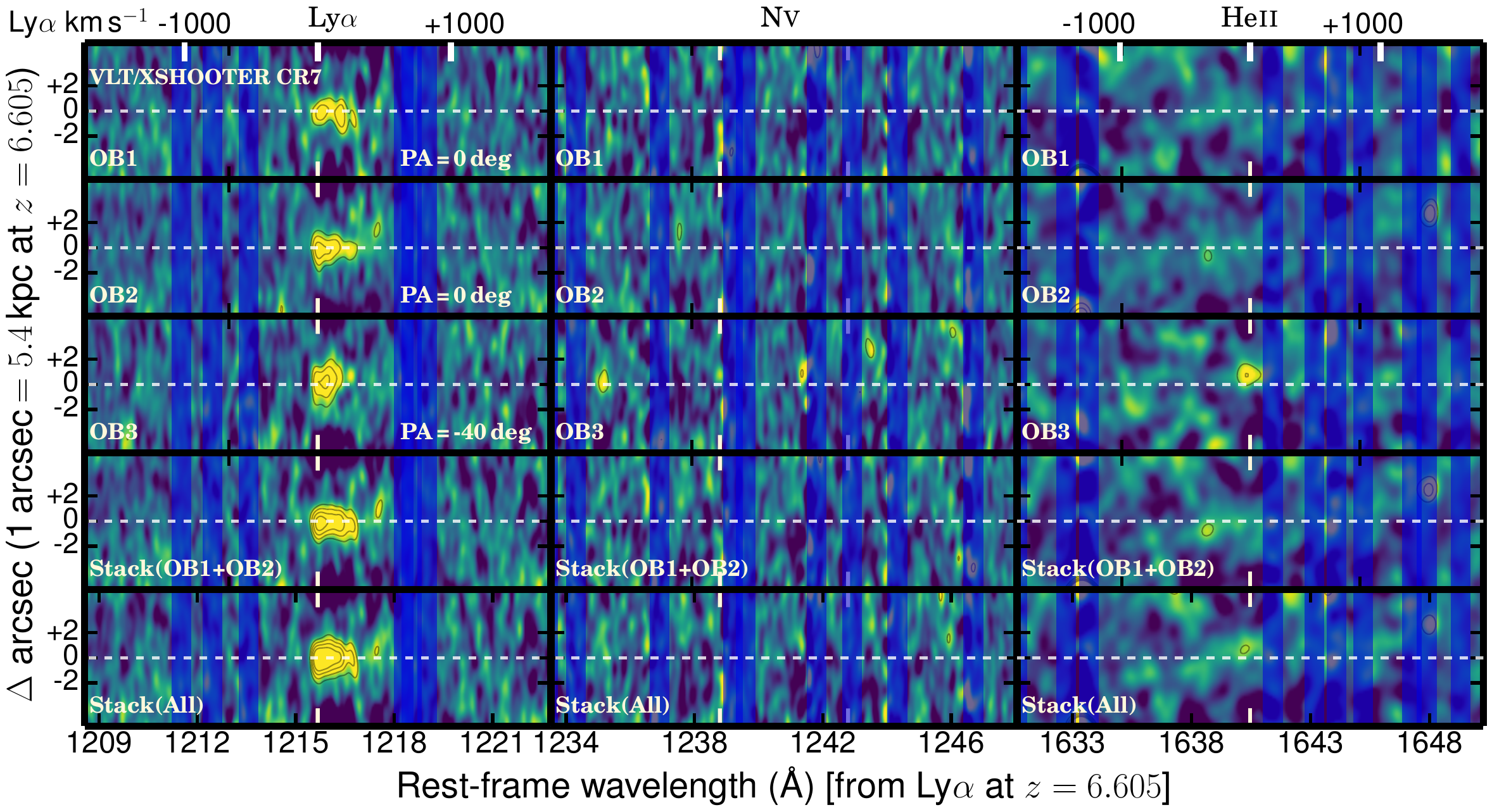}
\caption{Our final reduced 2D X-SHOOTER spectra, zoomed-in at the expected positions of Ly$\alpha$, N{\sc v} and He{\sc ii}. {We use a 3 spectral-spatial pixel Gaussian kernel to smooth the data and we show data in S/N space. Spatial contours show the 2, 3, 4 and 5 $\sigma$ levels and we use contrast cut-offs at $-1$ and $+2$\,$\sigma$. The location of sky lines are also labelled. He{\sc ii} is detected in OB3 at a $\approx3-4$\,$\sigma$ level (depending on the statistical method) with a spatial offset of $+0.8''$ towards clump B. In OB3 we also find a tentative emission line blue-shifted by $\sim800-900$\,km\,s$^{-1}$ to the expected wavelength of N{\sc v}, but we find that this is $<2.5$\,$\sigma$ in our analysis and thus not significant with the current data}.}
\label{Xshooter_2D_per_OB}
\end{figure*}

{When we analyse OB1 and OB2 separately (see Figure \ref{XSHOOTER_1D_HeII}), or when we stack these without OB3 we find no significant evidence of the presence of He{\sc ii} above $2.5$\,$\sigma$. For the stack of OB1 and OB2, sampling a PA angle of 0 degrees, we find a He{\sc ii} flux upper limit (2.5\,$\sigma$) of $<4.1\times10^{-17}$\,erg\,s$^{-1}$ (Table \ref{XSHOOT_MC}). However, stacking the three different OBs together leads to a detection of He{\sc ii} at the $\approx3.3\,\sigma$ level in our analysis, with a flux of $2.0^{+0.6}_{-0.6}\times10^{-17}$\,erg\,s$^{-1}$\,cm$^{-2}$. The lower flux we find compared to \citetalias{Sobral2015} is due to the different flux calibration which in \citetalias{Sobral2015} was based on UltraVISTA $J$ band. Finally, in Figure \ref{XSHOOTER_1D_MCMC_HeII} we show the results of our MC analysis for OB3 which contain the observations that dominate the He{\sc ii} signal. We compare it to the results presented in \citetalias{Sobral2015} after correcting them for the wavelength offset (see e.g. Figure \ref{OH_offset_lines} and \S\ref{xshoot_obs}), converting $\lambda_{\rm air}$ to $\lambda_{\rm vacuum}$ and scaling the counts to flux. We find a general good agreement within our errors, consistent with the signal being dominated by OB3. Note that in our analysis we do not smooth the data or bin it beyond 1/3 of the resolution, unlike \citetalias{Sobral2015}.

While we recover the He{\sc ii} emission line and identify the signal as coming from OB3 we still measure a lower significance than reported in \citetalias{Sobral2015}. This is mostly driven by the different methods used here, together with a new reduction. Furthermore, in order to place such reduced significance of an emission-line at high redshift into context \citep[see also][]{Shibuya2017_spec}, we investigate spectra of $z\sim6-8$ sources with published detections of high ionisation UV lines in the literature. We find that in general lines are less statistically significant or, in some cases, consistent with not being detected above $2.5$\,$\sigma$ in our framework. For example, we recover results for COSz2 \citep[][]{Laporte2017}, there is partial agreement for COSY \citep[][]{Stark2017,Smit2017,Laporte2017}, but we fail to detect ($<2.5$\,$\sigma$) Ly$\alpha$ for A2744 \citep[][]{Laporte_2017ALMA}. We present a more general comparison and discussion between our MC analysis and more widely used methods in the literature to measure the S/N of lines in Appendix \ref{Comparison_MC}.}

\subsubsection{Searching for other lines in X-SHOOTER}

We conduct an investigation of the full X-SHOOTER spectra, both on the full stack and also per OB. We search for UV rest-frame lines with FWHMs from 150 to 1500\,km\,s$^{-1}$ with redshifts from $z=6.58$ to $z=6.606$. {In addition, we also follow the methodology of \cite{Sobral2018b}. We do not detect any line above 2.5\,$\sigma$ apart from Ly$\alpha$ and He{\sc ii}. We nevertheless note that there could be a potential emission line below 2.5\,$\sigma$ in OB3. We find it in the VIS arm (showing the negatives from offsetting along the slit; see Figure \ref{Xshooter_2D_per_OB}) spatially coincident with Ly$\alpha$. For $z=6.60$ the potential emission line (S/N$\sim2$) is closest to the expected rest-frame wavelength of the N{\sc v} doublet (see Figure \ref{Xshooter_2D_per_OB}), but would imply a redshift of $z=6.583\pm0.001$ for it to be 1238.8\,{\AA} \citep[see e.g.][for N{\sc v} detections in other sources at $z\sim7$]{Tilvi2016,Hu2017,Laporte2017}.

\begin{figure}
\includegraphics[width=8.6cm]{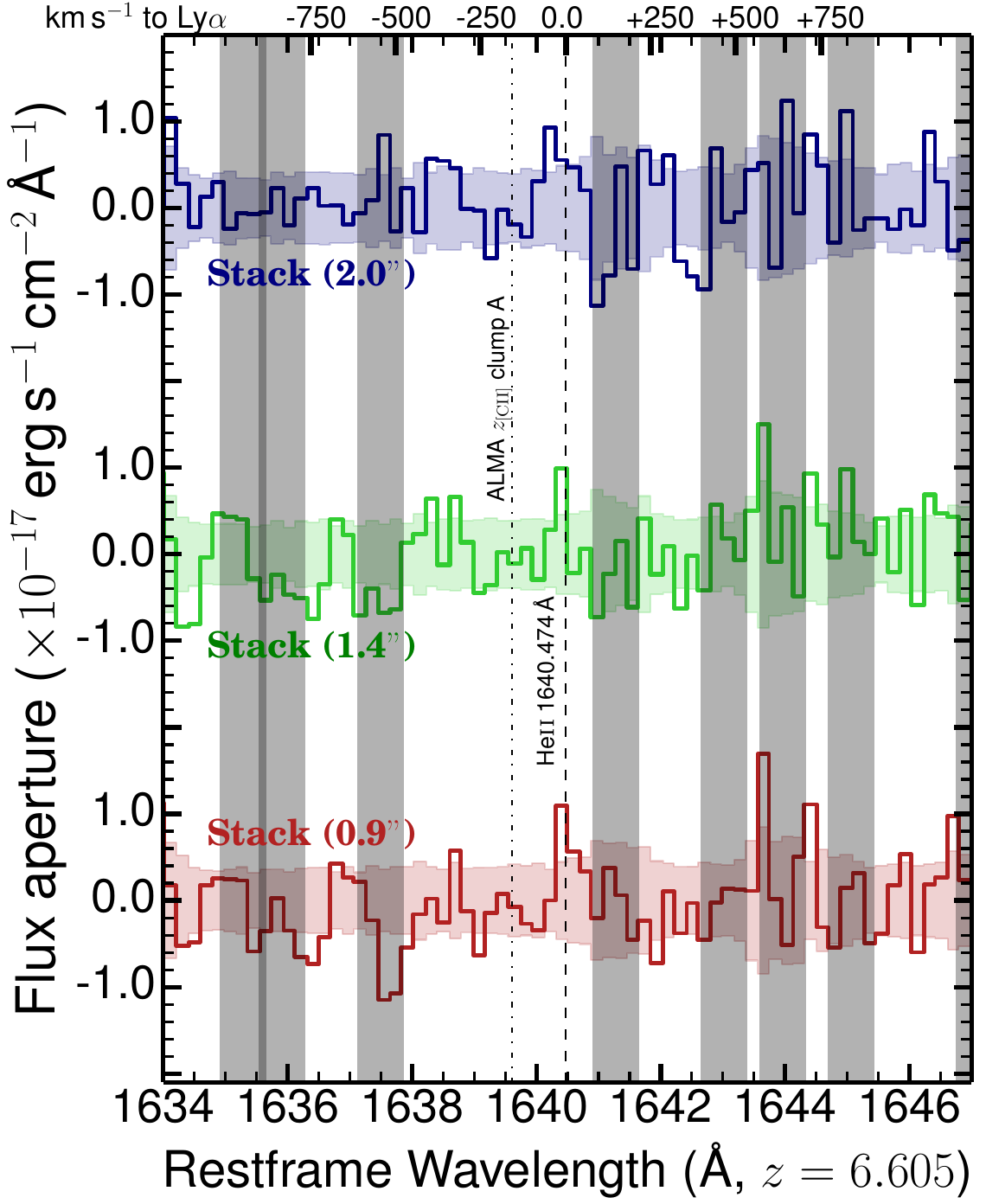}
\caption{The extracted 1D SINFONI spectra at the expected location of He{\sc ii} for stacks with different extraction apertures. The stacks show extractions obtained on the centre of the detector (assumed to trace the peak of Ly$\alpha$) using the appropriate aperture corrections based on the standard stars available. We conservatively estimate the noise with randomly placed apertures per wavelength slice per extraction. Sky lines are clearly labelled. We find a tentative line consistent with the same wavelength ($\lambda_{\rm vacuum,obs}=12475.3$\,\AA) as found with X-SHOOTER, but implying a lower flux close to $\approx0.5-1.0\times10^{-17}$\,erg\,s$^{-1}$\,cm$^{-2}$.}
\label{SINFONI_1D}
\end{figure}

\subsubsection{The nature of CR7 with SINFONI}\label{SINFONI_analysis}

One can further investigate the presence and flux of He{\sc ii} in CR7 by exploring SINFONI data. In Figure \ref{SINFONI_1D} we show the 1D stacks. We show these for different extraction apertures. We assume the source is in the centre of the 3D stacked cube which should correspond to the peak of Ly$\alpha$ emission due to the blind offset applied, per OB (see Section \ref{sinfoni_obs}). We visually search for potential emission in 2D by binning the data spectrally based on the He{\sc ii} signal in X-SHOOTER's OB3, and find a potential signal from He{\sc ii} in three of the OBs, with the strongest signal being found in the second OB, consistent with that found in \citetalias{Sobral2015} by using SINFONI data only. However, by measuring the noise on such wavelength slices (with apertures of $\sim1''$) we find that such signals on their own are of low significance ($<2$\,$\sigma$).

{Our MC analysis on the 1D stacks reveals tentative detections of He{\sc ii} at the $\approx2.5$\,$\sigma$ level for the 0.9$''$ and 1.4$''$ apertures (Figure \ref{SINFONI_1D}) used, yielding fluxes of $0.5^{+0.3}_{-0.2}\times10^{-18}$\,erg\,s$^{-1}$\,cm$^{-2}$ and a FWHM of $160\pm70$\,km\,s$^{-1}$. The line is found at a wavelength of $\lambda_{\rm vacuum,obs}=12475.3$\,{\AA}, matching very well the wavelength found with X-SHOOTER. If we use the $2.5$\,$\sigma$ as an upper limit for the He{\sc ii} flux assuming a non-detection we find $<1.3\times10^{-17}$\,erg\,s$^{-1}$\,cm$^{-2}$. This limit is consistent with the X-SHOOTER results, but favours a lower flux for He{\sc ii}, much closer to $\sim1\times10^{-17}$\,erg\,s$^{-1}$}. This would imply an observed He{\sc ii}/Ly$\alpha$ ratio of $\lsim0.06$. We find no other emission line in the SINFONI spectra for rest-frame wavelengths of $\sim1450-1770$\,{\AA}

\subsection{Variability: UltraVISTA}\label{Ultravista_variability}

We combine data from different epochs/data-releases of UltraVISTA \citep[][]{McCracken2012,Laigle2016} to constrain the potential variability of CR7. Note that CR7 is found very close to the overlap between the deeper/shallower UltraVISTA observations, with a strong gradient of exposure time and therefore depth in the East-West direction. We start by studying magnitudes obtained with different apertures and for mag-auto, contained in the public catalogue, both for $Y$ and $J$, tracking them from DR1 to DR2 and DR3. We find a large (in magnitude), $+0.51^{+0.14}_{-0.17}$\,mag variation\footnote{The magnitude difference is based on CR7 photometry, while errors are based on studying sources within 5\,arcmin of CR7; this allows to derive a more robust error which is higher than the formal error in the catalogue.} in the $J$ band mag-auto magnitude of CR7 from the UltraVISTA public catalogues from DR2 to DR3 \citep[see also][]{Bowler2017CR7}, while the magnitude stayed constant within the errors from DR1 to DR2 (see Appendix \ref{var_ultraVI_A}).

{In order to further investigate the potential variability of CR7 in the different data releases of UltraVISTA, we also conduct our own direct measurements on the data directly, fully available from the ESO archive. Furthermore, due to the potential problems with the usage of mag-auto, we use aperture photometry instead, placed over the UV clump A, at the centre of the CR7 system, and at the centre/peak of the Ly$\alpha$ emission: see Figure \ref{XSHOOTER_angle}. We measure AB magnitudes in apertures of 1.2$''$, 2$''$ and 3$''$ for $Y$, $J$, $H$ and $K$ and compare them with the measurements we obtain for DR2. For $H$ and $K$ the errors are always very large ($\approx0.5$\,mag) to investigate variability. Full details of our measurements are provided in Appendix \ref{var_ultraVI_A}.

Our results for aperture photometry on fixed positions for $Y$ and $J$ are presented in Figure \ref{Variability_UVista}. We find no significant changes/variability for any of the locations, apertures or bands, as all differences are $<2$\,$\sigma$. Similarly to \cite{Bowler2017CR7}, we find a change in the $J$ magnitude of CR7 in 2$''$ apertures of $0.21\pm0.12$ from DR2 to DR3 and in general there are weak trends of CR7 becoming fainter in fixed apertures from DR1 to DR3, but all these changes are at the $\sim1$\,$\sigma$ level. We therefore conclude that there is no convincing evidence for strong variability ($\Delta$\,mag\,$>0.3$) from the different DRs of UltraVISTA, but variability at the level of $\Delta$\,mag\,$\approx0.2$ is consistent with the data. }

\subsection{{\it HST} Grism observations: continuum results}

The spectrum of CR7 is extracted for its multiple UV components A, B and C detected with {\it HST} (see e.g. Figure \ref{XSHOOTER_angle}). We start by investigating the properties of the continuum and compare those with broad-band photometry. {We measure M$_{UV}$ (at rest-frame $\approx1500$\,\AA) by integrating the flux between rest-frame 1450\,{\AA} and 1550\,{\AA}, and also by fitting a power law of the form $\lambda^{\beta}$ between rest-frame 1450\,{\AA} and 2150\,{\AA}. All measurements are conducted per UV clump and by independently perturbing each spectral element within its Gaussian uncertainty and re-fitting 10,000 times. We present the median of all best fits, along with the 16th and 84th percentiles as the lower and upper errors in Table \ref{Best_fits_continuum}.} 

We find that our extraction of clump A yields {$\beta=-2.5^{+0.6}_{-0.7}$ and $M_{UV}=-21.87^{+0.25}_{-0.20}$}. Our results are consistent with the photometric properties of the clump estimated as {$\beta=-2.3\pm0.4$} and $M_{UV}=-21.6\pm0.1$ \citepalias[e.g.][]{Matthee2017_ALMA}, { although our measurement is completely independent of Ly$\alpha$ corrections which had to be applied in \citetalias{Matthee2017_ALMA} as F110W is contaminated by Ly$\alpha$ \citep[see also][]{Bowler2017CR7}}. This shows we are able to recover the continuum properties of clump A, and that these continuum properties show no significant evidence for variability within the errors.

{For the fainter clump B we find much more uncertain values of $\beta$ and $M_{UV}$ (see Table \ref{Best_fits_continuum}), consistent within the errors with $\beta=-1.0\pm1.0$ and $M_{UV}=-19.6\pm0.7$ from photometry \citepalias[see e.g.][]{Matthee2017_ALMA}. For clump C we do not make any significant continuum detection and we can only constrain $M_{UV}$ poorly.}

%
 %
\begin{table}
\centering
\caption{The rest-frame UV properties of the three UV clumps in CR7 constrained with {\it HST}/WFC3 grism data. $M_{\rm UV, integral}$ is estimated from integrating the spectrum directly between rest-frame 1450\,{\AA} and 1550\,{\AA}. We provide the best power-law fits: $\beta$ and the corresponding $M_{UV, \beta}$ computed as the value of the best fit at $\lambda_0=1500$\,\AA. Values for each measurement are the median of all best fits and the upper and lower errors are the 16th and 84th percentiles.}
 \label{Best_fits_continuum}
\begin{tabular}{cccccc}
\hline
Clump & M$_{\rm UV, integral}$ & $\beta$ & M$_{\rm UV, \beta}$ \\ 
\hline
A & $-21.87^{+0.25}_{-0.20}$ & $-2.5^{+0.6}_{-0.7}$  & $-22.02^{+0.14}_{-0.13}$ \\
B & $-21.0^{+0.5}_{-0.3}$ & $-2.6^{+1.7}_{-1.7}$  & $-20.9^{+0.4}_{-0.3}$ \\
C & $-20.2^{+0.8}_{-0.4}$ & -- & -- \\
\hline
\end{tabular}
\end{table}

%
%
\begin{table*}
 \centering
\caption{Results of our photometric study with {\it HST} data taken in 2012 and compared with more recent data taken with the same filters in 2017. We provide measurements centred on each clump and on the full system (see Figure \ref{XSHOOTER_angle}), both for apertures that capture each sub-component more optimally, but also with fixed $1''$ apertures. Errors are the 16th and 84th percentiles. We note that we do not apply corrections for the Ly$\alpha$ contribution to F110W. $\Delta$\,F110W, $\Delta$\,F160W and $\Delta\beta_{\rm UV}$ are computed using F110W and F160W photometry and differences between 2017 and 2012 observations. For further details, see Appendix \ref{var_HST}.}
 \label{VAR_HST_}
\begin{tabular}{cccccccc}
 \hline 
 Component  &  \multicolumn{2}{c}{2012-03-02}   &   \multicolumn{2}{c}{2017-03-14}   &    \multicolumn{3}{c}{$\Delta$: 2017 - 2012}   \\
 (Aperture) & F110W  &  F160W & F110W  &  F160W & $\Delta$\,F110W &  $\Delta$\,F160W &    $\Delta\beta_{\rm UV}$  \\
 \hline
 A (0.8$"$)  & $24.89^{+0.04}_{-0.04}$ & $25.07^{+0.07}_{-0.07}$ & $24.89^{+0.04}_{-0.04}$ & $24.96^{+0.07}_{-0.07}$  & $-0.01^{+0.06}_{-0.05}$ & $-0.12^{+0.10}_{-0.10}$  & $0.3^{+0.4}_{-0.4}$  \\
 B (0.4$"$)  & $27.04^{+0.15}_{-0.13}$ & $26.70^{+0.17}_{-0.15}$ & $26.99^{+0.13}_{-0.11}$ & $27.04^{+0.27}_{-0.22}$  & $-0.05^{+0.18}_{-0.19}$ & $0.33^{+0.30}_{-0.29}$  & $-1.2^{+1.1}_{-1.1}$  \\
 C (0.4$"$)  & $26.67^{+0.10}_{-0.09}$ & $26.51^{+0.14}_{-0.13}$ & $26.49^{+0.08}_{-0.08}$ & $26.29^{+0.13}_{-0.11}$  & $-0.18^{+0.12}_{-0.13}$ & $-0.23^{+0.17}_{-0.17}$  & $0.1^{+0.6}_{-0.7}$  \\
 CR7 (2.0$"$)  & $24.41^{+0.10}_{-0.08}$ & $24.24^{+0.08}_{-0.07}$ & $24.19^{+0.07}_{-0.05}$ & $24.36^{+0.13}_{-0.12}$  & $-0.22^{+0.10}_{-0.11}$ & $0.12^{+0.15}_{-0.15}$  & $-1.0^{+0.6}_{-0.6}$  \\
 CR7 (3.0$"$)  & $24.36^{+0.25}_{-0.17}$ & $24.11^{+0.10}_{-0.09}$ & $24.08^{+0.10}_{-0.07}$ & $24.27^{+0.26}_{-0.20}$  & $-0.28^{+0.19}_{-0.23}$ & $0.16^{+0.25}_{-0.23}$  & $-1.4^{+1.0}_{-1.1}$  \\
 \hline
  A (1.0$"$)  & $24.82^{+0.05}_{-0.05}$ & $24.97^{+0.08}_{-0.08}$ & $24.78^{+0.05}_{-0.04}$ & $24.91^{+0.09}_{-0.08}$  & $-0.04^{+0.06}_{-0.06}$ & $-0.06^{+0.11}_{-0.11}$ & $0.1^{+0.4}_{-0.4}$    \\
  B (1.0$"$)  & $26.53^{+0.49}_{-0.35}$ & $26.01^{+0.20}_{-0.18}$ & $26.05^{+0.15}_{-0.13}$ & $26.60^{+0.59}_{-0.41}$  & $-0.48^{+0.42}_{-0.46}$ & $0.58^{+0.59}_{-0.51}$ & $-3.3^{+2.1}_{-2.2}$    \\
  C (1.0$"$)  & $26.38^{+0.35}_{-0.26}$ & $25.80^{+0.16}_{-0.15}$ & $25.97^{+0.14}_{-0.11}$ & $25.79^{+0.21}_{-0.19}$  & $-0.41^{+0.31}_{-0.34}$ & $-0.02^{+0.25}_{-0.25}$ & $-1.2^{+1.2}_{-1.3}$    \\
  CR7 (1.0$"$)  & $25.63^{+0.11}_{-0.11}$ & $25.47^{+0.12}_{-0.11}$ & $25.47^{+0.09}_{-0.08}$ & $25.53^{+0.16}_{-0.15}$  & $-0.17^{+0.14}_{-0.14}$ & $0.06^{+0.20}_{-0.20}$ & $-0.7^{+0.8}_{-0.8}$    \\
  CR7 (1.0$"$)  & $25.63^{+0.11}_{-0.11}$ & $25.47^{+0.12}_{-0.11}$ & $25.47^{+0.09}_{-0.08}$ & $25.53^{+0.16}_{-0.15}$  & $-0.17^{+0.13}_{-0.14}$ & $0.06^{+0.20}_{-0.19}$ & $-0.7^{+0.7}_{-0.7}$    \\
\hline
\end{tabular}
\end{table*}

\subsection{{\it HST}/WFC3 imaging: is CR7 variable?}

Our grism detection of continuum in B (albeit at low S/N) and non-detection of C is perhaps unexpected given that previous UV photometry implied clump C was slightly brighter than B \citep[e.g.][]{Bowler2017CR7}. While our grism data is simply not constraining enough to investigate variability, new available imaging data taken in 2017 with WFC3 (program 14596, PI: Fan) with the same filters as in 2012 allow the opportunity to investigate variability in CR7 as a whole or in its individual components. The full details of our measurements are discussed in Appendix \ref{var_HST}.

\begin{figure}
\includegraphics[width=8.4cm]{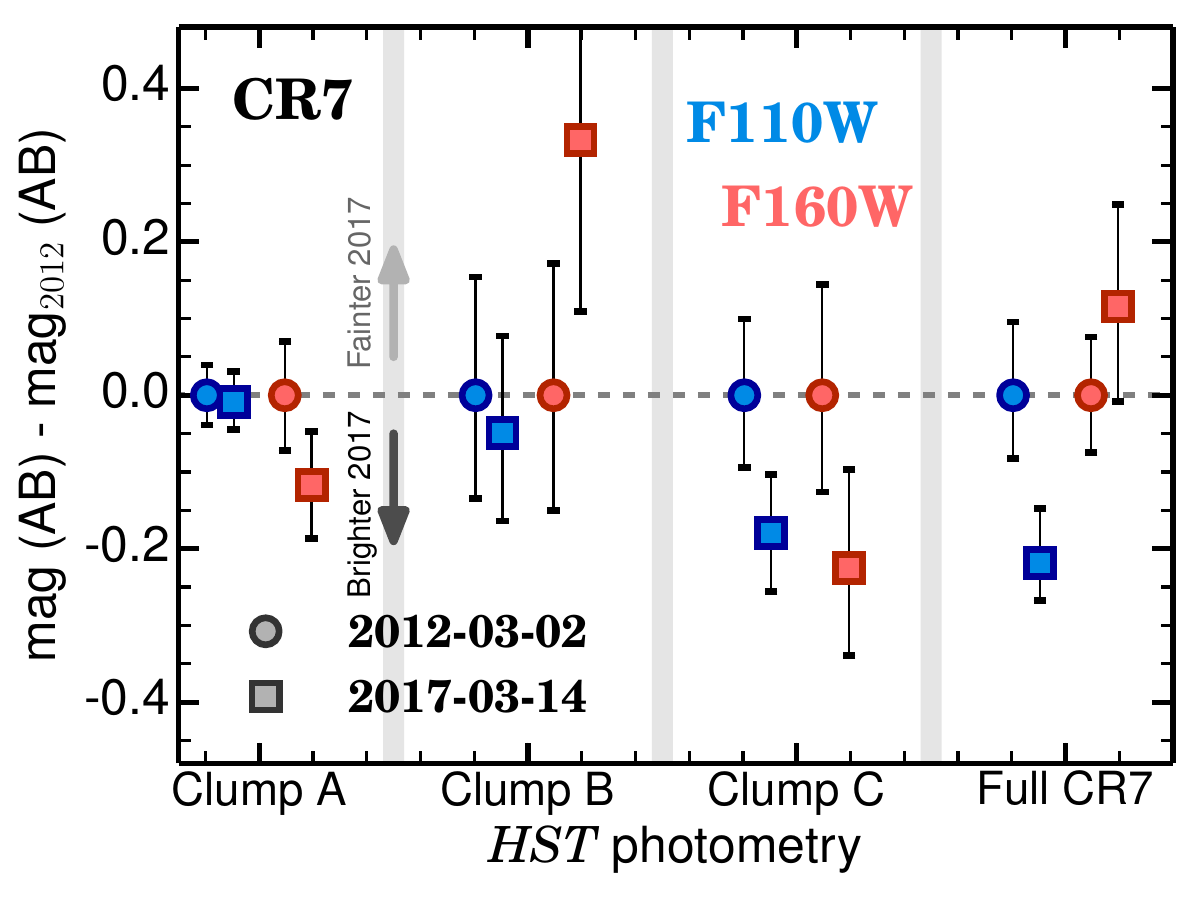}
\caption{{The difference in magnitudes for each UV clump in CR7, measured from {\it HST}/WFC3 photometry with the F110W and F160W filters in 2012 and in recent data taken in 2017. We find that while there is tentative evidence for clump C to have become brighter from 2012 to 2017 (when both bands are taken together), there is no convincing evidence for any of the clumps individually to have varied. However, the system as a whole is found to be brighter in the F110W filter by $-0.22^{+0.10}_{-0.11}$\,mag. We find this to be due to both clump C and inter-clump light, particularly between clumps C and B.}}
\label{Variability_HST}
\end{figure}

We present our results, obtained with apertures (diameter) of 0.8$''$, 0.4$''$ and 0.4$''$ placed on clumps A, B and C in Table \ref{VAR_HST_} and Figure \ref{Variability_HST}. We measure the full CR7 system, including any inter-clump UV light, with an aperture of 2$''$ (see Table \ref{VAR_HST_} for measurements with 1$''$ apertures centred on each component); see Figure \ref{XSHOOTER_angle}. The errors are estimated by placing apertures with the same size in multiple empty regions around the source and taking the 16th and 84th percentiles. As Figure \ref{Variability_HST} shows, there is no significant indication of variability for clumps A or B within the errors. The same is found for clump C in each individual band, although we find C to be brighter in 2017 by $\approx0.2$\,mag in both F110W and F160W, with the combined change providing some tentative evidence for variability. As a full system, CR7 became brighter by $0.22\pm0.10$\,mag, significant at just over $\approx2$\,$\sigma$. This brightening seems to be caused in part by clump C, but in addition to flux in between the UV clumps. Further observations taken even more recently with {\it HST}/WFC3 program 14596 (PI: Fan; not publicly available yet) will be able to further clarify/confirm our results.

%
%
\begin{figure*}
\includegraphics[width=15.cm]{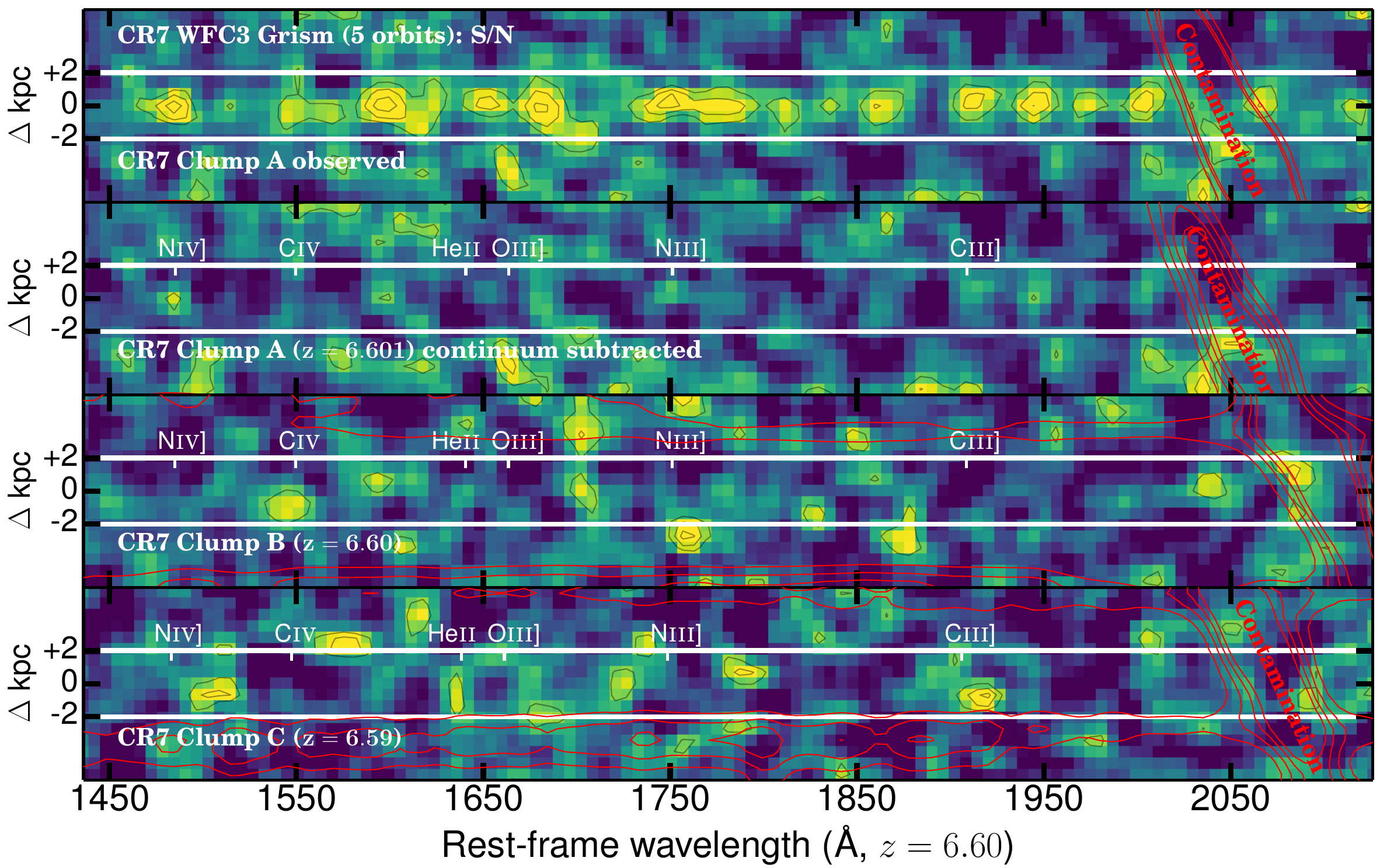}
\caption{The final {\it HST}/WFC3 Grism 2D reduced spectra, smoothed by 1 spatial-spectral pixel, for each of the three UV clumps in CR7: A, B and C (see Figure \ref{XSHOOTER_angle}). {All 2D here are shown in S/N space (contours: 2, 3, 4, 5\,$\sigma$), with the noise estimated away {from} the location where each clump is found. We use contrast cut-offs of $-1$\,$\sigma$ and $+3$\,$\sigma$.} For A, we show both the observed spectra (top) and the continuum subtracted 2D spectra. We show locations which were contaminated by nearby sources (contamination was subtracted but can still result in residuals). We also show the expected location of rest-frame UV lines using redshifts obtained with ALMA-[C{\sc ii}] \citepalias{Matthee2017_ALMA} close to the position of each clump and also an indicative ``slit" of 0.7$''$ that would contain close to 100\% of the flux of each clump. We note that our 1D extraction is based on the 2D image of {\it HST} of each clump. Apart from detecting continuum, no clear emission line $>3$\,$\sigma$ is found for any of the three clumps.}
\label{Grism2D}
\end{figure*}

\subsection{Grism observations: emission-line results}

{Figure \ref{Grism2D} presents the reduced {\it HST}/WFC3 2D spectra of each of the three clumps in CR7. For clump A we show both the observed (continuum-dominated) spectrum, along with the continuum subtracted, while for clumps B and C we show the observed spectrum only.} In Figure \ref{Grism_G140_ClumpsABC} we present the extracted 1D spectra of each clump.

%
%
\begin{table}
 \centering
\caption{Results from the MCMC chain to constrain the line fluxes of each clump within CR7 for our {\it HST/WFC3} grism data (A, B and C) after subtracting the UV continuum per clump. {We show the central value (best flux) and the percentiles, corresponding to $\pm1$\,$\sigma$ and $\pm2$\,$\sigma$.} All fluxes are in 10$^{-18}$\,erg\,s$^{-1}$\,cm$^{-2}$. We find no significant detection above 3\,$\sigma$ of any UV line within any of the clumps. However, we find potential detections of N{\sc iv}] in clump A and He{\sc ii} in clump C, both at over 2\,$\sigma$.}
 \label{Emcee}
\begin{tabular}{cccccc}
\hline
Emission &  2.5\% & 16\% & 50\% & 84\% & 97.5\% \\ 
Clump A  &  $-2\sigma$ & $-1\sigma$ & central & $+1\sigma$ & $+2\sigma$ \\ 
\hline
\bf N{\sc iv]}\,1485 &   0.21 &    7.66  &  14.78  &  20.94 &   26.69 \\
C{\sc iv}\,1549.5 &   -7.45  &  -2.23  &   3.13  &   8.70  &  15.07 \\
He{\sc ii}\,1640.5 &   -13.60  &  -8.70  &  -3.66  &   1.48  &   5.46 \\
O{\sc iii]}\,1663.5 &  -6.36  &  -2.27  &   2.33  &   7.11  &  11.22 \\
N{\sc iii]}\,1751 & -2.54  &   1.26  &   5.11  &   9.00  &  13.37 \\
C{\sc iii]}\,1908.5 &   -5.45  &  -2.36   &  1.22  &   4.76  &  7.49 \\
\hline
Clump B  &  $-2\sigma$ & $-1\sigma$ & central & $+1\sigma$ & $+2\sigma$ \\ 
\hline
N{\sc iv]}\,1485 &  -16.04  &  -11.06  &   -5.16  &    0.67  &    6.50 \\
C{\sc iv}\,1549.5 &   -3.72  &    0.60  &    5.00  &    9.93  &   15.20 \\
He{\sc ii}\,1640.5 &  -10.28  &   -6.35  &   -2.07  &    1.87  &    5.99 \\
O{\sc iii]}\,1663.5 &  -14.24 &   -10.49  &   -6.08  &   -2.11  &    2.22 \\
N{\sc iii]}\,1751 & -9.27 &    -5.67  &   -2.04  &    1.42  &    4.33 \\
C{\sc iii]}\,1908.5 &  -15.26 &   -12.33  &   -9.18  &   -6.14  &  -3.02 \\
\hline
Clump C  &  $-2\sigma$ & $-1\sigma$ & central & $+1\sigma$ & $+2\sigma$ \\ 
\hline
N{\sc iv]}\,1485 &  -7.29   &  -1.88   &   4.52   &  10.06   &  15.44 \\
C{\sc iv}\,1549.5 &  -10.57   &  -6.43   &  -2.34   &   2.24   &   6.72 \\
\bf He{\sc ii}\,1640.5 &  1.83   &   5.70   &   9.60   &  13.66   &  17.12 \\
O{\sc iii]}\,1663.5 &  -8.08   &  -3.87   &   0.23   &   4.45   &   8.01 \\
N{\sc iii]}\,1751 & -4.19   &  -0.92   &   2.42   &   5.58   &   8.62 \\
C{\sc iii]}\,1908.5 &  -1.93   &   0.98   &   4.01   &   7.09  &   9.83 \\
\hline
\end{tabular}
\end{table}

By using the best continuum fits shown in Figure \ref{Grism_G140_ClumpsABC}, we then continuum subtract the spectrum of each clump in order to look for any emission or absorption lines. We find no clear rest-frame UV emission or absorption line above a 3\,$\sigma$ level in any of the three clumps. Nonetheless, {there are tentative signals which are above $\sim2$\,$\sigma$}: N{\sc iv}] for the extraction of clump A ($z=6.60\pm0.01$) and He{\sc ii} for clump C (which would imply $z=6.58\pm0.01$). Note that while N{\sc iv}] \citep[see also][]{McGreer2017} for clump A is consistent with the systemic redshift now obtained for clump A with ALMA \citepalias{Matthee2017_ALMA}, the potential He{\sc ii} detection towards C would be consistent with a redshift of $z=6.58-6.59$. This could be related with the blue-shifted [C{\sc ii}] component found with ALMA towards C.

In order to better quantify the significance of all rest-frame UV lines, we measure all lines with {\sc Grizli}\footnote{https://github.com/gbrammer/grizli/}/Emcee (MCMC), {by fitting simultaneously to all of the exposure level 2D spectra, which is much more appropriate to grism data \citep[see e.g.][]{Kummel2009,Brammer2012,Momcheva2016}}. We obtain the 2.5, 16, 50, 84 and 97.5 percentiles of the Emcee chain, and show the results in Table \ref{Emcee}. {Our results show that there are no clear ($>3$\,$\sigma$) emission line detections in either of the UV clumps. We also obtain very strong constraints on He{\sc ii} centred on UV clumps A and B, showing no detections, with the 2\,$\sigma$ limit for He{\sc ii} flux in each of those clumps being $<6\times10^{-18}$\,erg\,s$^{-1}$\,cm$^{-2}$}. This strongly implies that any He{\sc ii} signal in X-SHOOTER is not coming directly from the UV components of either A or B, in agreement with the X-SHOOTER results, as otherwise it should have been detected at a $\sim4-5$\,$\sigma$ level. Interestingly, for clump C there is a potential signal from He{\sc ii} (see Table \ref{Emcee}), as we find that 97.5\% of realisations result in a He{\sc ii} flux of up to $17.1\times10^{-18}$\,erg\,s$^{-1}$\,cm$^{-2}$, with a central value of $(10\pm4)\times10^{-18}$\,erg\,s$^{-1}$\,cm$^{-2}$.

Furthermore, in order to conduct our full analysis self-consistently, we also apply our MC analysis in the same way as for X-SHOOTER and SINFONI \citep[][]{Sobral2018b} on the extracted 1D grism spectra per clump. We find that N{\sc iv}] in clump A and He{\sc ii} in clump C are significant at just above $2.5$\,$\sigma$, while all the other lines are $<2.5$\,$\sigma$. The full results, including the limits\footnote{ In order to estimate conservative $2.5$\,$\sigma$ limits in a self-consistent way we determine the $-2.5$\,$\sigma$ and $2.5$\,$\sigma$ flux values (corresponding to 0.62 and 99.38 percentiles) and shift the mid-point between both to a flux of zero as we assume a non-detection. Our $2.5$\,$\sigma$ upper limit is then determined as the difference between 0 and $2.5$\,$\sigma$.} for the lines that we do not detect above 2.5\,$\sigma$ are provided in Table \ref{table_lines}.
}

%
%
\begin{figure*}
\includegraphics[width=15.5cm]{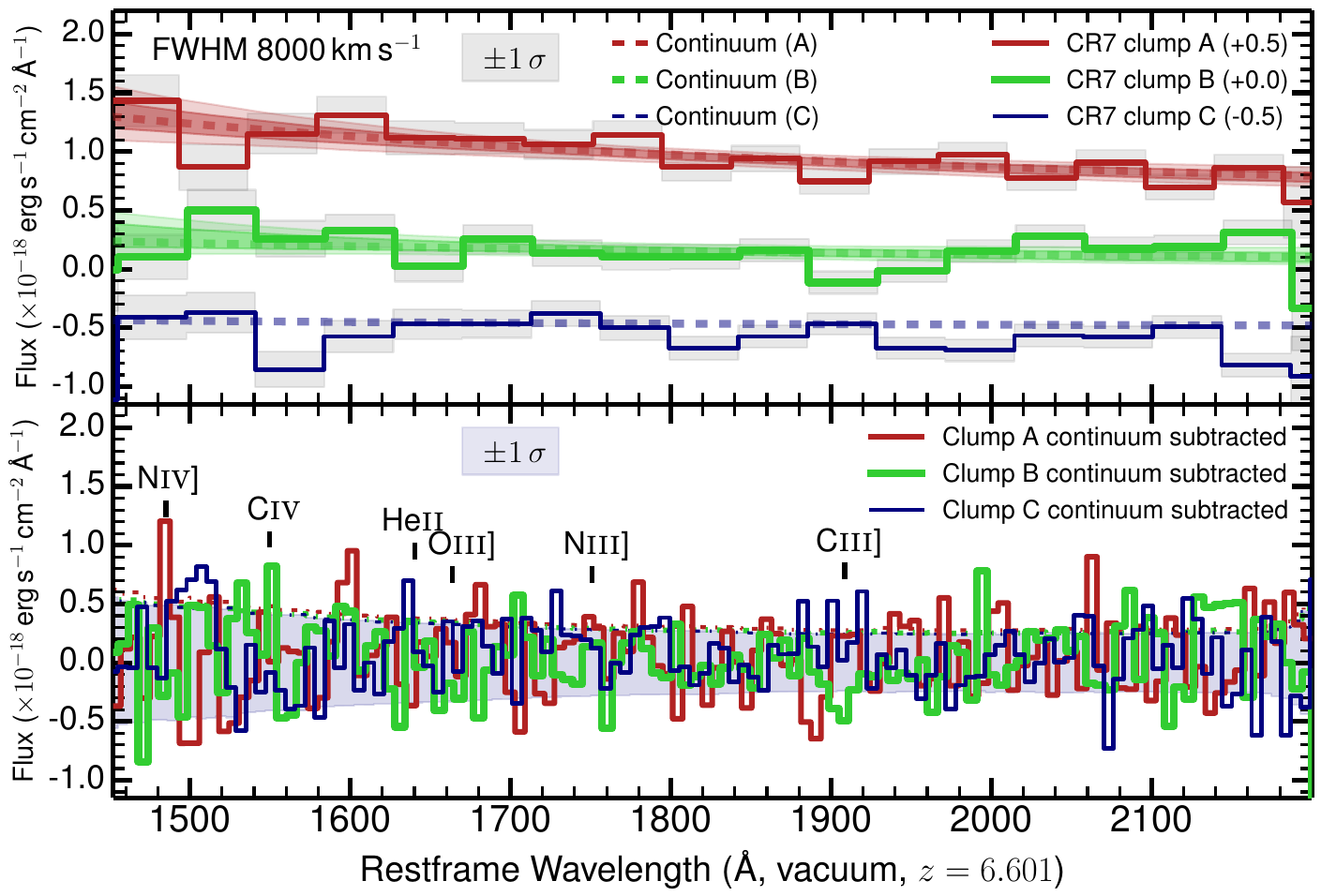}
\caption{{\it HST}/WFC3 grism 1D spectra of the {three UV clumps of CR7 extracted based on the UV detections of each clump in the pre- and post-images with the F140W filter}. {\it Top:} Clump A is significantly detected in the UV continuum and is well fitted with {$\beta=-2.5^{+0.6}_{-0.7}$ and $M_{UV}=-21.87^{+0.25}_{-0.20}$; we show the 16-84 and 2.3-97.7 percentile contours for all fits}. Clump B is also detected in the rest-frame continuum but at a much lower significance, while clump C is not significantly detected in the continuum. {\it Bottom:} After continuum subtracting the spectra of each clump we find no significant detection above 3\,$\sigma$ of any rest-frame UV line. There are only tentative detections of N{\sc iv}] in clump A and He{\sc ii} in clump C. The resolved spectra also show that any potential He{\sc ii} emission {from the UV clumps} would have to likely come from or near clump C and not clump A. We assign relatively strong limits to all observed rest-frame UV lines, which we use to further interpret CR7.}
\label{Grism_G140_ClumpsABC}
\end{figure*}

\section{CLOUDY modelling and the physical conditions of CR7}\label{JWST_Predictions}

Here we explore the best constraints on a variety of lines (see Tables \ref{table_lines} and \ref{table_lines_EW_limtis}) to infer the possible physical properties of CR7, exploring its uniqueness as a $z\sim7$ source for which we already have a wealth of resolved information despite the limited amount of telescope time {invested.}

%
%
\begin{table*}
{
 \centering
\caption{A summary of the high ionisation rest-frame UV lines investigated for CR7 and/or their upper limits constrained in this work {with our MC analysis. All fluxes are in units of 10$^{-17}$\,erg\,s$^{-1}$\,cm$^{-2}$}. We list and use them in vacuum. {We list fluxes for $\gsim2.5$\,$\sigma$ detections, or the $<2.5$\,$\sigma$ upper limits constraints for the ``full system" (X-SHOOTER and SINFONI) and also for each of the clumps A, B, C from the {\it HST}/WFC3 grism data}. $^*$Ly$\alpha$ flux from the X-SHOOTER slit (observed) implies $5.9\pm0.5\times10^{-17}$\,erg\,s$^{-1}$\,cm$^{-2}$ {(integration without Gaussian fitting). Ly$\alpha$ is clearly extended \citepalias{Sobral2015}, and thus the full slit losses are larger than for a simple point-source; the full Ly$\alpha$ flux over the full CR7 system is estimated as $17\pm1$ \,erg\,s$^{-1}$\,cm$^{-2}$ \citep[see][]{Matthee2017}. We follow \citetalias{Matthee2017_ALMA} and associate Ly$\alpha$ observed fluxes to clumps A, B and C based on the 2D Ly$\alpha$ distribution from NB91 photometry. Note that these Ly$\alpha$ fluxes have not been measured directly with spectroscopy, and are thus very uncertain}.}

 \label{table_lines}
\begin{tabular}{ccccccc}
\hline
 Emission line & Ionisation  & CR7 (Slit)  & CR7 (0.9$''$) & Clump A  & Clump B & Clump C  \\ 
$\lambda_{\rm vacuum}$ (\AA)  & Energy (eV) & X-SHOOTER & SINFONI & WFC3 & WFC3  & WFC3 \\ 
\hline
Ly$\alpha$\,1215.67 & 13.6 & $17\pm1$* & --- & $8.3\pm0.7^*$ & $2.7\pm0.5^*$ & $1.3\pm0.4^*$  \\ 
N{\sc v}\,1238.8,1242.78  &  77.4  &  $<1.4$  &   ---   & ---   & ---   & ---   \\ 
O{\sc iv]}\,1401,1407  &  54.9  &  $<3.0$  &   ---   & ---   & ---   & ---   \\ 
N{\sc iv]}\,1483.4,1486.6  &  47.4  &  $<2.2$  &  $<5.1$ & $1.9^{+0.7}_{-0.7}$  & $<2.6$ & $<2.8$  \\ 
C{\sc iv}\,1548.2,1550.77  &  47.9  &  $<1.5$  &  $<1.0$ & $<2.3$ & $<2.2$ & $<2.1$  \\ 
He{\sc ii}\,1640.47  &  54.4  &  $2.0^{+0.6}_{-0.6}$  &  $0.5^{+0.3}_{-0.2}$ & $<2.7$ & $<1.9$ & $1.1^{+0.5}_{-0.4}$   \\ 
O{\sc iii]}\,1661,1666  &  35.1  &  $<2.6$  &  $<2.6$ & $<2.3$ & $<1.8$ & $<1.8$  \\ 
N{\sc iii]}\,1749.7,1752.2  &  29.6  &  $<15.9$  &  $<30.7$ & $<1.6$ & $<1.5$ & $<1.7$  \\ 
C{\sc iii]}\,1907,1910  &  24.4  &  $<1.7$  &   ---  & $<1.5$ & $<1.3$ & $<1.4$  \\

\hline
\end{tabular}
}
\end{table*}

In order to explore a relatively wide range of physical conditions that may be found in CR7, we use the {\sc cloudy} (v 13.03) photo-ionisation code \citep{CLOUDY1998,CLOUDY2013}. {Further details are given in \cite{Sobral2018b}}. Table \ref{CLOUDY_table} summarises the key physical conditions. Briefly, we use three kinds of models \citep[for a similar, more extensive analysis, see also, e.g.][]{Nakajima2017}: i) power-laws to mimic the spectra of AGN, ii) stellar spectra from {\sc bpass} \citep[][]{Eldridge2009,Eldridge2017,Stanway2016} and iii) black body models to further interpret and make simple predictions. We note that as a first step, and for simplicity, we only ionise the gas using photons. Shock ionisation may in principle also play a role \citep[e.g.][]{Allen2008,Jaskot2016}, which could be explored once observations provide detections in a range of lines, and particularly to explore spatially resolved emission-line ratio maps \citep[see e.g.][]{Miley2008,Morais2017,Comerford2017}.

\subsection{The physical conditions in CR7 with current constraints: the full system}

We use our simple {\sc cloudy} grid predictions and the methodology presented in \cite{Sobral2018b} to interpret what the current measurements and constraints of several lines in CR7 imply. {We start by investigating the ``full" CR7 system as a whole using flux measurements from X-SHOOTER and SINFONI. We note that if one assumes that no line is detected apart from Ly$\alpha$ and only upper limits are used, models are, not surprisingly, completely unconstrained.}

%
%
\begin{table}
 \centering
\caption{Rest-frame EW$_0$ {constraints} for UV lines (see Table \ref{table_lines}) inferred from our WFC3 grism observations of each of the three UV clumps of CR7. As a comparison, we also provide equivalent measurements for the ``full" CR7 system based on our X-SHOOTER flux constraints. We use the flux limits provided in Table \ref{table_lines} and [$M_{UV},\beta$] of [$-22.2\pm0.1,-2.2\pm0.4$], [$-22.0\pm0.2,-2.5\pm0.7$], [$-20.9\pm0.4,-2.6\pm1.7$] and [$-20.1\pm0.3,-2.3\pm0.8$] for the full system, clumps A, B and C, respectively, in order to predict the continuum at the rest-frame wavelength of each emission line. We list fluxes accompanied by the 16 and 84 percentiles if a line is significant at$\gsim2.5$\,$\sigma$ or we list the $<2.5$\,$\sigma$ non-detections constraints.}
 \label{table_lines_EW_limtis}
\begin{tabular}{ccccc}
\hline
Emission & CR7 & Clump A  & Clump B & Clump C  \\ 
Line & (\AA) & (\AA) &  (\AA) &  (\AA)  \\ 
\hline
Ly$\alpha$  &   $122^{+16}_{-15}$  &  $65^{+16}_{-12}$ & $57^{+45}_{-24}$  & $63^{+28}_{-26}$ \\ 
N{\sc v}  &   $<11$  &  ---  & ---  & --- \\ 
N{\sc iv]}  &   $<24$  &  $24^{+11}_{-9}$ & $<96$  & $<214$ \\ 
C{\sc iv}  &   $<18$  &  $<34$  & $<88$  & $<180$ \\ 
He{\sc ii}  &   $26^{+9}_{-9}$  &  $<45$ & $<88$  & $98^{+49}_{-43}$ \\ 
O{\sc iii]}  &   $<37$  &  $<39$  & $<85$  & $<182$ \\ 
N{\sc iii]}  &   $<253$  &  $<32$  & $<85$  & $<195$ \\ 
C{\sc iii]}  &   $<32$  &  $<37$  & $<95$  & $<189$ \\ 
\hline
\end{tabular}

\end{table}

{Due to the He{\sc ii} flux constraints for the full system as a whole (implying a rest-frame EW of $26\pm9$\,{\AA}; see Tables \ref{table_lines} and \ref{table_lines_EW_limtis}), we find that standard {\sc bpass} models at ``normal" metallicities struggle to fully reproduce some of the observations, although, as \cite{Bowler2017CR7} showed, modified {\sc bpass} models with super-solar $\alpha$ elements at extremely low metallicity are able to reproduce the observations \citep[see][]{Bowler2017CR7}. Furthermore, our simple power-law and black body models can both easily reproduce the observations, implying gas-phase metallicities of $\approx10-20$\% solar and ionisation parameters of $\log\,U\approx-3$, but with large uncertainties of over 1\,dex in all parameters using our very wide model grid.}

\subsection{CR7 resolved: the nature of each individual UV clump} \label{physics_clumps}

{For clump A, the tentative detection of N{\sc iv}] (with an EW$_0$ of $24^{+11}_{-9}$\,{\AA}) and the non-detections of other lines, allow to place some constraints on the nature of the source, suggesting a high Nitrogen abundance and a high effective temperature, closer to $T_{\rm eff}\sim100$\,K. However, there is currently no strong evidence for the presence of an AGN, as stellar models (particularly at lower metallicities and/or with binaries) can reproduce the emission line ratios within the large uncertainties. Nevertheless, the metallicity is consistent with $\approx0.1-0.2$\,$Z_{\odot}$ as suggested byALMA observations (based on the [C{\sc ii}]/UV ratio; \citetalias{Matthee2017_ALMA}).

Current flux and EW upper limits for clump B (Tables \ref{table_lines} and \ref{table_lines_EW_limtis}) do not allow to truly constrain the physical conditions that we explore, but we note that ALMA results hint for a metallicity of $\approx0.1-0.2$\,$Z_{\odot}$. Our non-detection of any high ionisation UV lines in clump B does not provide any evidence for an unusually high ionisation parameter or for strong AGN activity \citep[see also e.g.][]{Nakajima2017}, although some AGN activity is still possible. We further constrain the physical conditions using the UV+FIR SFR measured per clump \citepalias{Matthee2017_ALMA}, not allowing models to significantly over or underestimate by factors of more than two the SFR per clump.

For clump C, the tentative detection of He{\sc ii} at a very high EW (with EW$_0$ of $98^{+49}_{-43}$\,{\AA})} brings in some evidence of its potential AGN nature, while the non-detections of the other lines are also consistent with a potential low metallicity AGN. By using all constraints, models suggest that C can be powered by an ionisation source with roughly $\log\,U\approx-2$ and surrounded by a relatively low metallicity gas ($\approx0.1-0.2$\,$Z_{\odot}$), {but the constraints are currently very weak and deeper observations are required to improve the constraints; see e.g. Table \ref{table_lines_EW_limtis} \citep[see also][]{Dors2018}.}

We conclude that with the current uncertainties, all three clumps are consistent with being relatively young star-bursts with similar {metal-poor} gas-phase metallicities of $\sim0.05-0.2$\,Z$_{\odot}$. There is currently no strong evidence for the presence of an AGN in either clumps A or B, and there is only tentative evidence for clump C to have a higher ionisation parameter and to potentially host an AGN.

\section{Discussion}\label{Discussion}

CR7 has previously been discussed as being powered by very low metallicity stars (PopIII-like; \citealt{Sobral2015,Visbal2016}), as a candidate for being a DCBH \citep[e.g.][]{Sobral2015,Pallottini2015,Hartwig2016,Agarwal2016,Agarwal2017A}, or as hosting a significant population of young, binary stars and/or WR stars at extremely low metallicities \citep[e.g.][]{Bowler2017CR7}. The observed Ly$\alpha$ and He{\sc ii} EWs based on {UltraVISTA DR2 public} photometry in \citetalias{Sobral2015} could only be explained by an extreme hard ionising spectrum, implying a high effective temperature and an extremely low metallicity of $\approx0.05-0.5$\,\% solar \citep[][]{Hartwig2016,Bowler2017CR7}. Different components of CR7 have now been spectroscopically confirmed to be part of the same system \citepalias{Matthee2017_ALMA}, with velocity offsets of only a few hundred km\,s$^{-1}$ at most, and with evidence of dynamics/potential merging activity {(see Figure \ref{Summary_figure_CR7})}. New observations of CR7 reveal the unique potential of bright enough targets at high redshift, allowing the first spatially resolved studies of both rest-frame UV lines and [C{\sc ii}] detections with ALMA \citepalias{Matthee2017_ALMA}; {see also \cite{Carniani2017b}}.

%
%
\begin{figure*}
\includegraphics[width=14cm]{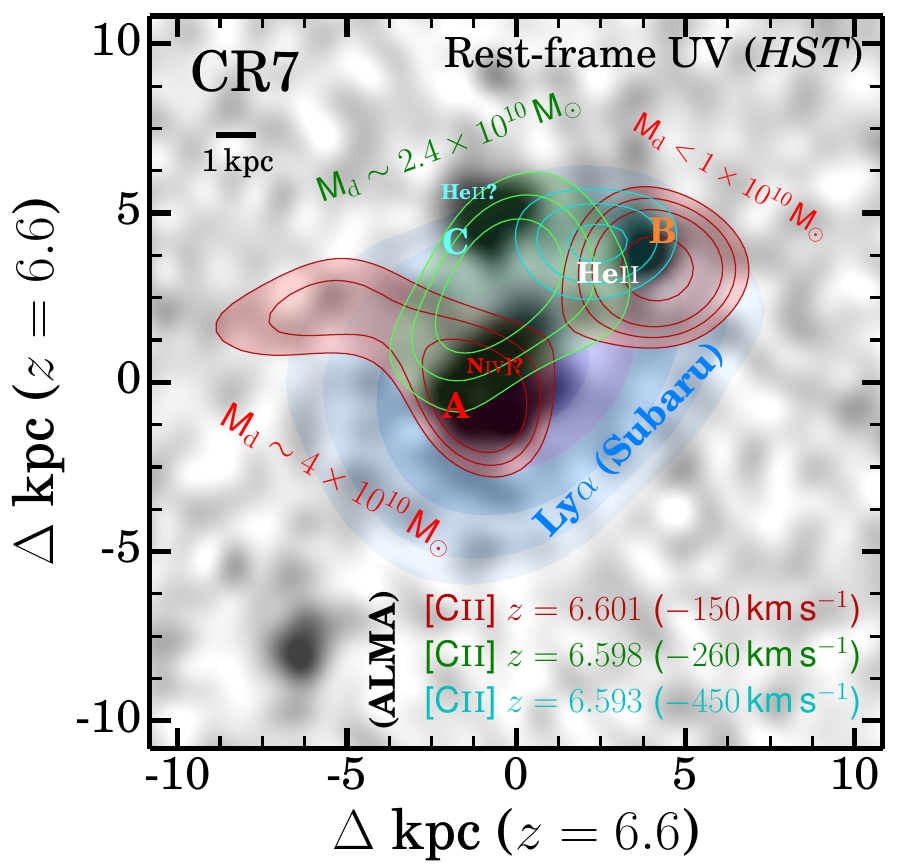}
\caption{A summary of the results presented or discussed in this paper on CR7, resulting from observations obtained with Subaru/S-Cam \citepalias{Sobral2015}, {\it HST}/WFC3 \citepalias[][and this study]{Sobral2015}, ALMA \citepalias{Matthee2017_ALMA} and the VLT (this study). We show how the Ly$\alpha$ halo extends over the 3 rest-frame UV clumps, peaking closer to the brightest clump A. [C{\sc ii}] follow-up with ALMA reveals at least 4 different components, with the brightest and likely most massive being coincident with clump A but extending beyond it. There is a blue-shifted component closest to clump C (but not coincident with it) which is potentially massive and close to the inferred location of He{\sc ii} found with X-SHOOTER in OB3 (labelled in white; see Figure \ref{XSHOOTER_angle}). We also indicate the tentative emission lines found in clumps A and C, but we note that those are $<3$\,$\sigma$.}
\label{Summary_figure_CR7}
\end{figure*}

Overall, and specifically for clump A, our results show that the He{\sc ii}/Ly$\alpha$ ratio is significantly lower than measured using UltraVISTA flux estimate of He{\sc ii} \citepalias{Sobral2015}, with this ratio being {more likely} below $\sim0.06$ instead of close to $\sim0.2$ {(see Sections \ref{XSHOT_HeII} and \ref{SINFONI_analysis})}. This rules out the most extreme DCBH scenarios for clump A. Together with the [C{\sc ii}] detection in A \citepalias{Matthee2017_ALMA}, our results imply a metallicity of roughly 0.1-0.2\,Z$_{\odot}$ for clump A (to be confirmed/verified with {\it JWST}), thus becoming globally inconsistent with a ``PopIII-like'' scenario metallicity \citep[$\sim0.005$\,Z$_\odot$;][]{Bowler2017CR7}. Our latest results indicate that A is a more `normal' starburst, consistent with feedback processes already fully in place, as indicated from the Ly$\alpha$ line profile modelling \citep[][]{Dijkstra_G_DS2016}. It is interesting that while ALMA provides a detection of Carbon \citepalias{Matthee2017_ALMA} in CR7's clump A (also in/around clumps B and C; {see Figure \ref{Summary_figure_CR7})}, and even though we estimate a metallicity of roughly 0.1-0.2\,Z$_{\odot}$ \citep[similar to sources studied by e.g.][]{StarkCIV}, we do not detect any high ionisation Carbon line (e.g. C{\sc iv} or C{\sc iii}]), down to rest-frame EW upper limits of $\approx37,34$\,{\AA} in C{\sc iii}] and C{\sc iv}, respectively. This is consistent with the hypothesis explored in \cite{Matthee2017} that current C{\sc iv} and C{\sc iii}] detections in galaxies at the epoch of re-ionisation are only possible for even {intrinsically} brighter sources with much higher SFRs (UV brighter or with significant lensing amplifications) and/or {AGN \citep[e.g.][]{Laporte2017,Shibuya2017_spec,Sobral2018b}}.

{Current observational constraints point towards CR7's clump C \citep[e.g.][]{Dors2018} or additional inter-clump components (Figure \ref{Summary_figure_CR7}) as the most puzzling and uncertain at the moment}. This component seems to show the largest blueshift and presents evidence for the presence of a high EW He{\sc ii} line (see Table \ref{table_lines_EW_limtis}). Furthermore, He{\sc ii} is also tentatively detected {between clumps B and C. While there are} indications that C may host an AGN/high ionisation UV source, this would be somewhat puzzling in other regards as it would imply a likely low black hole mass given its sub-L$^*$ luminosity in both observed Ly$\alpha$ and in the UV (in our grism observations it is the faintest component in CR7). {Some obscuration could in principle be invoked to explain the low luminosity in Ly$\alpha$ and the UV for clump C, but ALMA observations \citepalias{Matthee2017_ALMA} do not detect any dust. However, current ALMA observations are not sensitive to significantly hot dust that may be present in C.}

In principle, future X-ray observations may also help to determine the nature of these high redshift sources, but these may have to achieve significantly high resolution (if they are to locate AGN within multi-component galaxies) and be relatively deep to detect the presence of an AGN in e.g. clump C. Given its luminosity in the UV and also its potential He{\sc ii} luminosities, one would expect X-ray luminosities of $\approx10^{42}$\,erg\,s$^{-1}$, about $\sim4$ times lower than predicted by \cite{Pallottini2015} due to the much lower He{\sc ii} luminosity in clump C than originally {estimated using} UltraVISTA photometry \citepalias{Sobral2015}. Therefore, identifying AGN will likely be much more efficient with {\it JWST}, particularly with the IFU on NIRSpec, {at least until the launch of the {\it Athena} X-ray mission.}

Our results also point towards potential consequences when interpreting emission lines from clump C and from other clumps/the full system. Given the geometry of the system and the small velocity {offsets between components \citepalias[see][and Figure \ref{Summary_figure_CR7}]{Matthee2017_ALMA}, it is possible that each clump is differentially illuminated by a time-dependent AGN+SF composite SED}. Observations with {\it JWST} obtained over $\sim1-2$ years could be crucial to test how important any time-variability and the illumination from different clumps may be. This would be important to e.g. understand whether illumination from another clump (e.g. C) could give rise/be responsible for potential high ionisation lines {seen in the gas of another}. Until then, detailed 3D simulations could be performed with full radiation transfer in order to further investigate similar systems {\citep[e.g.][]{Carniani2017b,Matthee2018_COLA}} and allow to make specific predictions, not only for CR7, but for other similar sources within the epoch of re-ionisation, {prior to the launch of {\it JWST} in a few years}.


\section{Conclusions}\label{Conclusion}

We presented new {\it HST}/WFC3 grism {and imagining observations} and combined those with a re-analysis of {flux calibrated X-SHOOTER and SINFONI data obtained with the VLT for the most luminous Ly$\alpha$ emitter at $z=6.6$, COSMOS Redshift 7 (CR7; \citetalias{Sobral2015}). We investigated the continuum, variability and rest-frame UV lines of the source as a whole and its 3 UV components}. We find that:

\begin{itemize}

 \item {The Ly$\alpha$ profile of CR7 is broader in the East-West direction (FWHM\,$=300^{+56}_{-40}$\,km\,s$^{-1}$) compared to a PA angle of $-40$\,deg that matches the major axis of Ly$\alpha$ emission (FWHM\,$=180^{+50}_{-30}$\,km\,s$^{-1}$) and that connects clumps A and B. The stack of all OBs yields a Ly$\alpha$ FWHM\,$=270^{+35}_{-30}$\,km\,s$^{-1}$, in good agreement with \citetalias{Sobral2015}.}

 \item {Our re-reduced, flux calibrated X-SHOOTER stacked spectrum of CR7 reveals a $\approx 3\,\sigma$ He{\sc ii} detection in CR7 with a flux of $2.0^{+0.6}_{-0.6}\times10^{-17}$\,erg\,s$^{-1}$\,cm$^{-2}$. Such signal is found to be dominated by OB3 which on its own yields a flux of $3.4^{+1.0}_{-0.9}\times10^{-17}$\,erg\,s$^{-1}$\,cm$^{-2}$.}
 
\item {The He{\sc ii} line detected in OB3 is spatially offset by $+0.8''$ from clump A towards B but does not coincide with the UV clump B. The stack of OB1 and OB2 result in a non-detection ($<2.5$\,$\sigma$) of He{\sc ii}. {\it HST} grism data confirms that there is no strong He{\sc ii} emission directly on UV clumps A or B. Our re-reduced SINFONI data presents some evidence for He{\sc ii} but suggests a flux closer to $\approx1\times10^{-17}$\,erg\,s$^{-1}$\,cm$^{-2}$, with our MC method yielding $0.5^{+0.3}_{-0.2}\times10^{-17}$\,erg\,s$^{-1}$\,cm$^{-2}$.}

 \item {No statistically significant changes are seen in $Y$ photometry from DR2 to DR3, but we find a change of $+0.51^{+0.14}_{-0.17}$\,mag (mag-auto) in the UltraVISTA $J$ band public catalogue for CR7 as a whole from DR2 to DR3. However, we find no statistically significant variation ($<2$\,$\sigma$) when we conduct aperture photometry with carefully estimated errors.}

\item Our WFC3 grism spectra provide a significant detection of the UV continuum of CR7's clump A, yielding an excellent fit to a power law with {$\beta=-2.5^{+0.6}_{-0.7}$ and $M_{UV}=-21.87^{+0.25}_{-0.20}$}. This is fully consistent with the broad band photometry and with no variability for clump A.

 \item {Careful measurements of F110W and F160W data of CR7 taken in 2012 and 2017 reveal no significant variability in either bands for clumps A or B, but there is a tentative combined brightening of clump C. CR7 as a whole (aperture of 2$''$ encompassing the 3 clumps) changes by $-0.22^{+0.10}_{-0.11}$ in F110W, providing 2.2\,$\sigma$ evidence for variability. We find that this change can be explained by both clump C and also inter-clump light, but requires confirmation. No variability is seen in F110W in A (within $\pm0.05$\,mag) or B (within $\pm0.2$\,mag).}

\item  {{\it HST} grism data do not detect any rest-frame UV line in any of the UV clumps above 3\,$\sigma$, with rest-frame EW$_0$ limits varying from $<30$\,{\AA} to $<200$\,{\AA}. We find a tentative ($\approx2.5$\,$\sigma$) He{\sc ii} line in clump C's data, yielding a flux of $1.10^{+0.50}_{-0.46}\times10^{-17}$\,erg\,s$^{-1}$\,cm$^{-2}$ and $z=6.574^{+0.019}_{-0.013}$.}

\item Our results show that the He{\sc ii}/Ly$\alpha$ ratio {for clump A is significantly lower than measured using the UltraVISTA} flux estimate of He{\sc ii} \citepalias{Sobral2015}, with this ratio {being likely closer to$\lsim0.06$} instead of close to $\sim0.2$. This rules out the most extreme DCBH scenarios for clump A.

\item We perform CLOUDY modelling and obtain limits on the metallicity and constrain the ionising nature of CR7. We conclude that CR7 is likely actively forming stars without any clear AGN activity in clumps A and B, with a metallicity of $\sim0.1-0.2$\,Z$_{\odot}$ (to be confirmed/verified with {\it JWST}) and with component A experiencing the most massive starburst. Together with the [C{\sc ii}] detection in clumps A and B \citepalias{Matthee2017_ALMA}, our results are globally inconsistent with a ``PopIII-like'' scenario metallicity \citep[$\sim0.005$\,Z$_\odot$;][]{Bowler2017CR7} for clumps A and B.

\item {Component C, or an inter-clump component, may host a high ionisation source/AGN and could be variable, although the evidence for variability is only at the $\approx2.2$\,$\sigma$ level and requires further, deeper observations with {\it HST} to be confirmed.}

\end{itemize}

{Overall, our results reveal that CR7 is a complex system (see Figure \ref{Summary_figure_CR7}) which may be giving us an early glimpse of the complicated rapid assembly processes taking place in the early Universe. The high resolution observations presented here, those obtained with ALMA \citep[e.g.][]{Jones2017,Matthee2017_ALMA,Carniani2017b} and recent simulations for galaxies at $z~\sim7$ \citep[e.g.][]{Gallerani2016,Pallottini2017,Behrens2018} point towards early galaxies being chaotic collections of metal-poor merging clumps which will also likely bring along black holes and potentially lead to measurable variability}. Such complex systems imply that the approach of simply placing a very narrow slit in a single UV light peak may only reveal part of the full picture, particularly if there is significant ionising flux from nearby sources. {It seems that the systems studied so far at $z\sim7-8$ require spatially resolved observations, ideally obtained by IFU spectrographs, in order to identify the nature of different components \citep[e.g.][]{Carniani2017,Carniani2017b,Hashimoto2018}. The current results also reveal the importance of simulations to take into account} such complex systems by performing a full 3D radiation transfer for systems like CR7 and comparing with observations, particularly to constrain the role of multiple ionising sources. Until {\it JWST} is launched, further spatially resolved observations of {other bright enough systems which have been spectroscopically confirmed \citep[e.g.][]{Ouchi2013,Sobral2015,Hu2016,Matthee2017,Matthee2018_COLA,Carniani2018} with MUSE, ALMA and {\it HST} will assure an even more efficient and diverse laboratory to advance our knowledge of the early assembly of galaxies within the epoch of re-ionisation. These can then be further applied to fainter and more numerous sources.}

\section*{Acknowledgments}

{We thank the anonymous reviewer for the numerous detailed comments that led us to greatly improve the quality, extent and statistical robustness of this work}. DS acknowledges financial support from the Netherlands Organisation for Scientific research (NWO) through a Veni fellowship. JM acknowledges the support of a Huygens PhD fellowship from Leiden University. AF acknowledges support from the ERC Advanced Grant INTERSTELLAR H2020/740120. BD acknowledges financial support from NASA through the Astrophysics Data Analysis Program (ADAP), grant number NNX12AE20G and the National Science Foundation, grant number 1716907. {We are thankful for several discussions and constructive comments from Johannes Zabl, Eros Vanzella, Bo Milvang-Jensen, Henry McCracken, Max Gronke, Mark Dijkstra, Richard Ellis and Nicolas Laporte. We also thank Umar Burhanudin and Izzy Garland for taking part in the XGAL internship in Lancaster and for exploring the {\it HST} grism data independently.}

Based on observations obtained with {\it HST}/WFC3 {programs 12578, 14495 and 14596}. Based on observations of the National Japanese Observatory with the Suprime-Cam on the Subaru telescope (S14A-086) on the big island of Hawaii. This work is based in part on data products produced at Terapix available at the Canadian Astronomy Data Centre as part of the Canada-France-Hawaii Telescope Legacy Survey, a collaborative project of NRC and CNRS. Based on data products from observations made with ESO Telescopes at the La Silla Paranal Observatory under ESO programme IDs 294.A-5018, 294.A-5039, 092.A-0786, 093.A-0561, 097.A-0043, 097.A-0943, 098.A-0819, 298.A-5012 and 179.A-2005, and on data products produced by TERAPIX and the Cambridge Astronomy Survey Unit on behalf of the UltraVISTA consortium. The authors acknowledge the award of service time (SW2014b20) on the William Herschel Telescope (WHT). WHT and its service programme are operated on the island of La Palma by the Isaac Newton Group in the Spanish Observatorio del Roque de los Muchachos of the Instituto de Astrofisica de Canarias. This research was supported by the Munich Institute for Astro- and Particle Physics (MIAPP) of the DFG cluster of excellence ``Origin and Structure of the Universe".

We have benefited immensely from the public available programming language {\sc Python}, including {\sc NumPy} \& {\sc SciPy} \citep[][]{van2011numpy,jones}, {\sc Matplotlib} \citep[][]{Hunter:2007}, {\sc Astropy} \citep[][]{Astropy2013} and the {\sc Topcat} analysis program \citep{Topcat}. This research has made use of the VizieR catalogue access tool, CDS, Strasbourg, France.

All data used for this paper are publicly available, and we make all reduced data available with the refereed paper.


\bibliographystyle{mnras}
\bibliography{bib_LAEevo.bib}


\appendix

\section{Reduced data: public release}

We publicly release all spectroscopic and imaging data described and analysed in this paper. This includes the 2Ds from X-SHOOTER and {\it HST}/WFC3. We also release our extracted 1D spectra, flux calibrated, including our best estimate of the 1\,$\sigma$ noise per wavelength element. We release these as fits files, available to download with the refereed paper. Raw data are also publicly available for all the data-sets described here by querying the appropriate archives and proposal IDs.

\subsection{Comparison with \citetalias{Sobral2015}: the NIR wavelength calibration offset}

{In Figure \ref{OH_offset_lines} we show the offset between the wavelength calibration in the NIR from \citetalias{Sobral2015} and our reduction, resulting from the use of old arcs in \citetalias{Sobral2015}; applying an offset of $\approx+6.9$\,{\AA} to the 1D of \citetalias{Sobral2015} is able to correct the wavelengths in the range covering He{\sc ii}; see \S\ref{xshoot_obs}.}

%
\begin{figure}
\includegraphics[width=8.4cm]{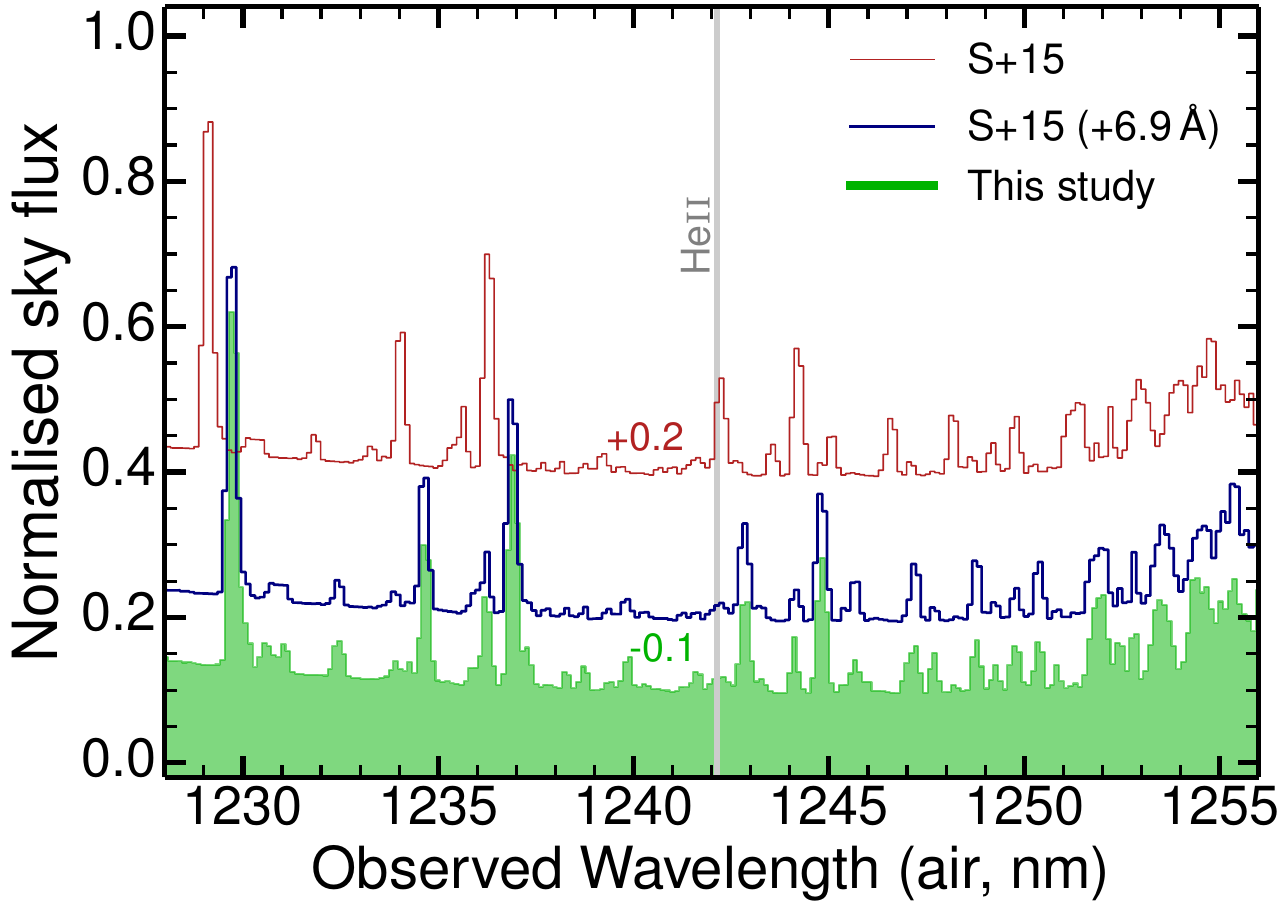}
\caption{The arbitrarily normalised sky spectrum for the stack of the three OBs in the NIR arm around the observed emission line identified as He{\sc ii} for CR7 for our reduction and a comparison to \citetalias{Sobral2015}, showing an offset in the wavelength calibration. Applying an offset of $+6.9$\,{\AA} to the NIR spectrum presented in \citetalias{Sobral2015} results in a good agreement with our results. Note that we have shifted the normalised sky spectra in the Y direction as indicated for clarity.}
\label{OH_offset_lines}
\end{figure}

\section{{\sc cloudy} modelling}\label{CLOUDY}

{We present the main parameters explored in our {\sc cloudy} modelling in Table \ref{CLOUDY_table}, and also release all the models/{\sc cloudy} grids in {\sc fits} format together with this paper. For more details, see \cite{Sobral2018b}.}

%
%
\begin{table}
{
 \centering
  \caption{Parameters and ranges used for the photo-ionisation {\sc cloudy} \citep{CLOUDY1998,CLOUDY2013} modelling presented in \citet{Sobral2018b} and used in this study. We vary density, metallicity and the ionisation parameter ($\log\,U$) for star-like ionisation, here modelled with BPASS \citep[][]{Eldridge2009,Stanway2016}, or more simply with black bodies of varying temperature from 20 to 160K. AGN-like ionisation is modelled using power-law sources with varying slopes.}
  \begin{tabular}{cc}
  \hline
   \bf Parameter & \bf Range used for all models   \\
 \hline
   \noalign{\smallskip}
  Density (n$_{\rm H}$\,cm$^{-3}$) & 30, 100, 300, 1000  \\
  Metallicity ($\rm \log\,Z_{\odot}$)  &  $-$2 to $+$0.5 (steps of 0.05)  \\ 
  Ion. param. ($\log\,U$) & $-$5 to $+$2 (steps of 0.2) \\
 \hline
  \bf Type of model & \bf Range used   \\
  \hline
   Black body (Temp., K) & 20k to 160k (steps of 1k) \\
   Power-law (slope) & $-$2.0 to $-$1.0 (steps of 0.05)  \\
   BPASS ($\log$\,Age, yr) & 6.0 to 8.9 (steps of 0.1) \\
  \hline 
\end{tabular}
\label{CLOUDY_table}
}
\end{table}

\section{Variability in UltraVISTA}\label{var_ultraVI_A}

\subsection{Public catalogues}

{In order to understand the flux differences in the $J$ band for CR7 for different public UltraVISTA data \citep[][]{McCracken2012} releases (DRs), we check how the magnitude of CR7 has changed in the 3 DRs of the UltraVISTA survey. We retrieve public catalogues from the ESO archive and include all sources that are i) detected in all UltraVISTA data releases and ii) are within 5\,arcmin separation from CR7. CR7 itself is only detected in all three releases in the $Y$ and $J$ bands and thus we focus on these bands. DR1 was released in February 2012, while DR2 was released in January 2014 and DR3 in April 2016. While DR1 has an average exposure time of $\sim50$\,ks (including the deep stripes), the DR2 exposure time at the location of CR7 is 46\,ks. DR3 does not seem to have added any exposure time to the region where CR7 is found, with DR3 listing a total exposure time of 44.6\,ks, down from 46\,ks in DR2. We use aperture photometry in $1''$, $2''$ and mag-auto provided in the public catalogues and show the results in Figures \ref{fig:DR2_DR3_apertures} and \ref{fig:DR2_DR3}. The quadrature combined photometric error of DR2 and DR3 would imply that the change in mag-auto for CR7 is statistically significant at a $\approx4.6$\,$\sigma$ level. However, we find a few other sources in the public catalogue for which such change in magnitude also happens. Furthermore, we derive a more conservative error, based on the 16th and 84th percentiles of all magnitude changes between DR2 and DR3 for sources in the vicinity of CR7, yielding a change from DR2 to DR3 of $+0.51^{+0.14}_{-0.17}$\,mag in $J$ mag-auto. Using the public catalogue, this implies a 3\,$\sigma$ statistical significance for the $J$ band. However, as Figure \ref{fig:DR2_DR3_apertures} shows, the mag-auto variation seems to be the most extreme, and variations in the $J$ band for aperture photometry are less significant, except for 1$''$.

A variation of $+0.51^{+0.14}_{-0.17}$\,mag (becoming fainter) in the $J$ band is a significant change in the public catalogue ($\approx3$\,$\sigma$), and dramatically affects the interpretation of a high EW emission line in the J band \citetalias[which was taken as a strong prior in][]{Sobral2015}. However, we caution that even though we use conservative errors based on the public catalogue (the formal errors would imply a change closer to 5\,$\sigma$), we find that there are a few other sources with a similar magnitude change in the vicinity of CR7 (see Figure \ref{fig:DR2_DR3}). We are therefore cautious in interpreting this change in magnitude as intrinsic variability of CR7 using the public catalogues. For the $Y$ band for example, we find no evidence for variability within the 1\,$\sigma$ uncertainties. For more detailed results, see Figures \ref{fig:DR2_DR3_apertures} and \ref{fig:DR2_DR3}.}

%
%
\begin{figure}
\includegraphics[width=8.6cm]{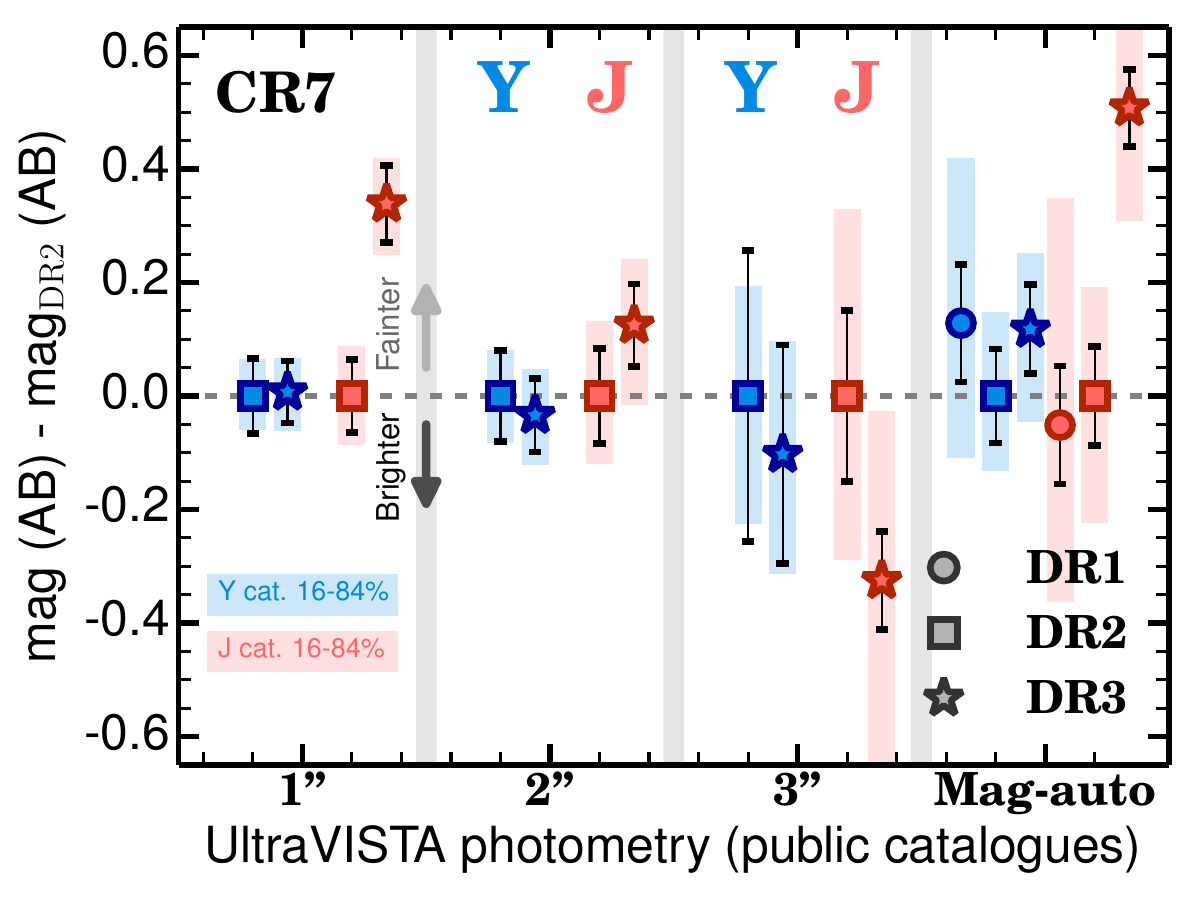}
\caption{{Comparison between $Y$ and $J$ AB magnitudes in public UltraVISTA catalogues for DR1, DR2 and DR3. We study the potential variation in relation to DR2 (used in \citetalias{Sobral2015}) for magnitudes measured with apertures (diameter) of 1$''$, 2$''$, 3$''$ and mag-auto. We show the errors provided in the public catalogues, but we also estimate more conservative errors by computing the 16th and 84th percentiles of the change in magnitude from one DR to the next of sources in the vicinity of CR7 with magnitudes between 23 and 25. We find no statistical significant variation in $Y$. The variations in $J$ from DR2 to DR3 in both 1$''$ and mag-auto are above 2\,$\sigma$.}}
\label{fig:DR2_DR3_apertures}
\end{figure}

\subsection{Public images/data}

{We use the ESO archive to obtain the reduced DR1, DR2 and DR3 UltraVISTA mosaics in $Y$, $J$, $H$ and $K$. We make cut-outs of all images centred on CR7 and assure that they are well aligned. We perform aperture photometry on the positions defined in Figure \ref{XSHOOTER_angle} for all bands and for all data releases. In order to estimate the error, we use {\sc sextractor} \citep[][]{Bertin1996} to produce a segmentation map and place 1,000 apertures with the same size in empty regions in the image and then compute the 16th and 84th percentiles as our errors. We also compute the median of all empty apertures and subtract it before computing the flux or magnitude for a given aperture, in order to subtract the local sky/background. We note that CR7 is in the transition between shallow and deeper UltraVISTA data. Due to this, we concentrate our analysis in a region of $\approx30''\times30''$ and measure the local noise in this region. In order to correct our aperture photometry measurements we follow \cite{Bowler2017CR7} and apply the necessary corrections\footnote{We assume that our apertures of $1.2''$, $2.0''$ and $3.0''$ in [$Y$, $J$, $H$, $K$] recover [0.56,0.6,0.63,0.64], [0.79,0.83,0.85,0.87] and [0.92,0.94,0.95,0.96] of the total flux, respectively.}. 

}

%
%
\begin{figure}
\includegraphics[width=8.6cm]{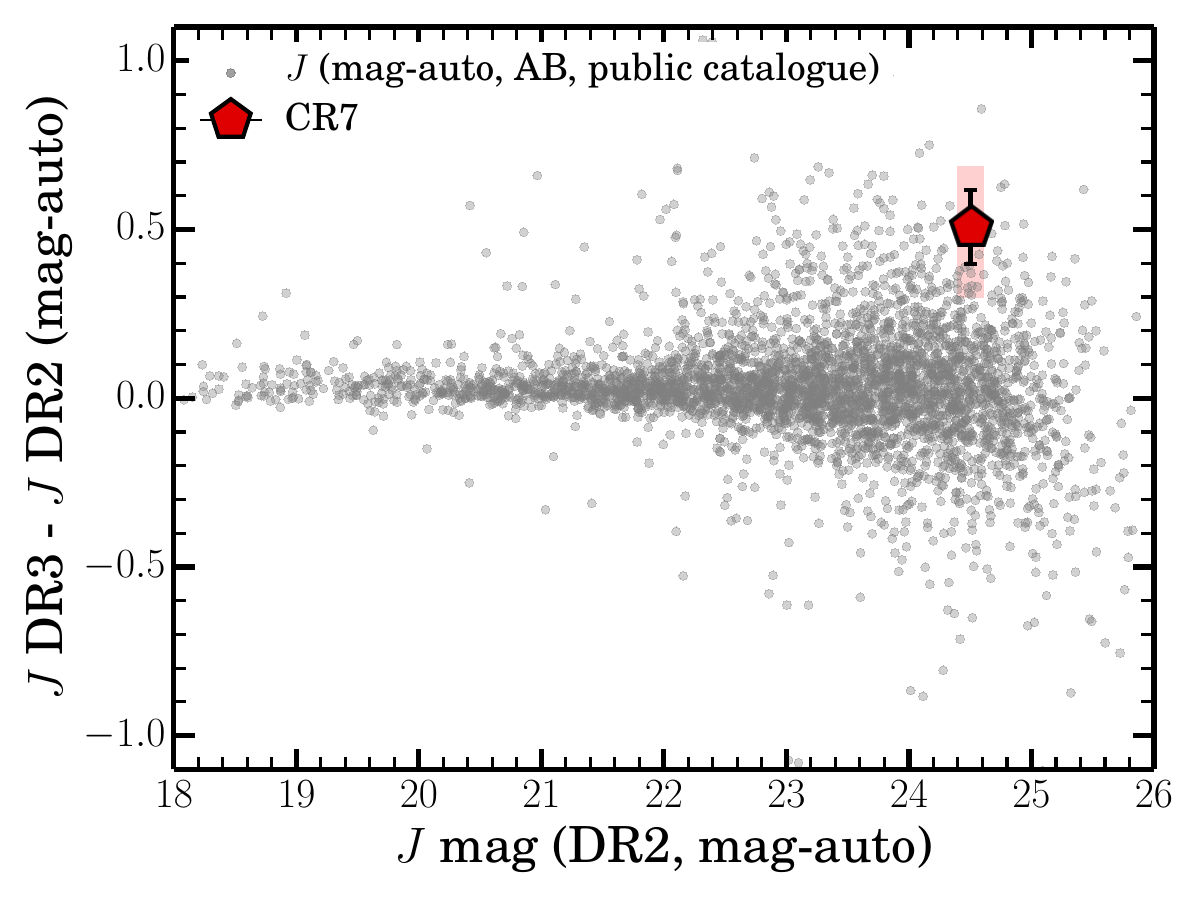}
\caption{{Comparison between mag-auto magnitudes in the public  DR3 and DR2 UltraVISTA $J$ catalogues. We show CR7 and also all matched sources between DR2 and DR3 that are within 5\,arcmin of CR7. The quadrature combined photometric error of DR2 and DR3 implies that the change in mag-auto for CR7 is statistically significant by $\approx4.6$\,$\sigma$, but we note that there are a few other sources for which this change in magnitude also happens. Motivated by this, we derive a more conservative error, based on the 16th and 84th percentiles of all magnitude changes between DR2 and DR3 for sources near CR7, yielding a change from DR2 to DR3 of $+0.51^{+0.14}_{-0.17}$\,mag, suggesting a 3\,$\sigma$ variation, based on the public catalogues.}}
\label{fig:DR2_DR3}
\end{figure}

\begin{figure}
\includegraphics[width=8.4cm]{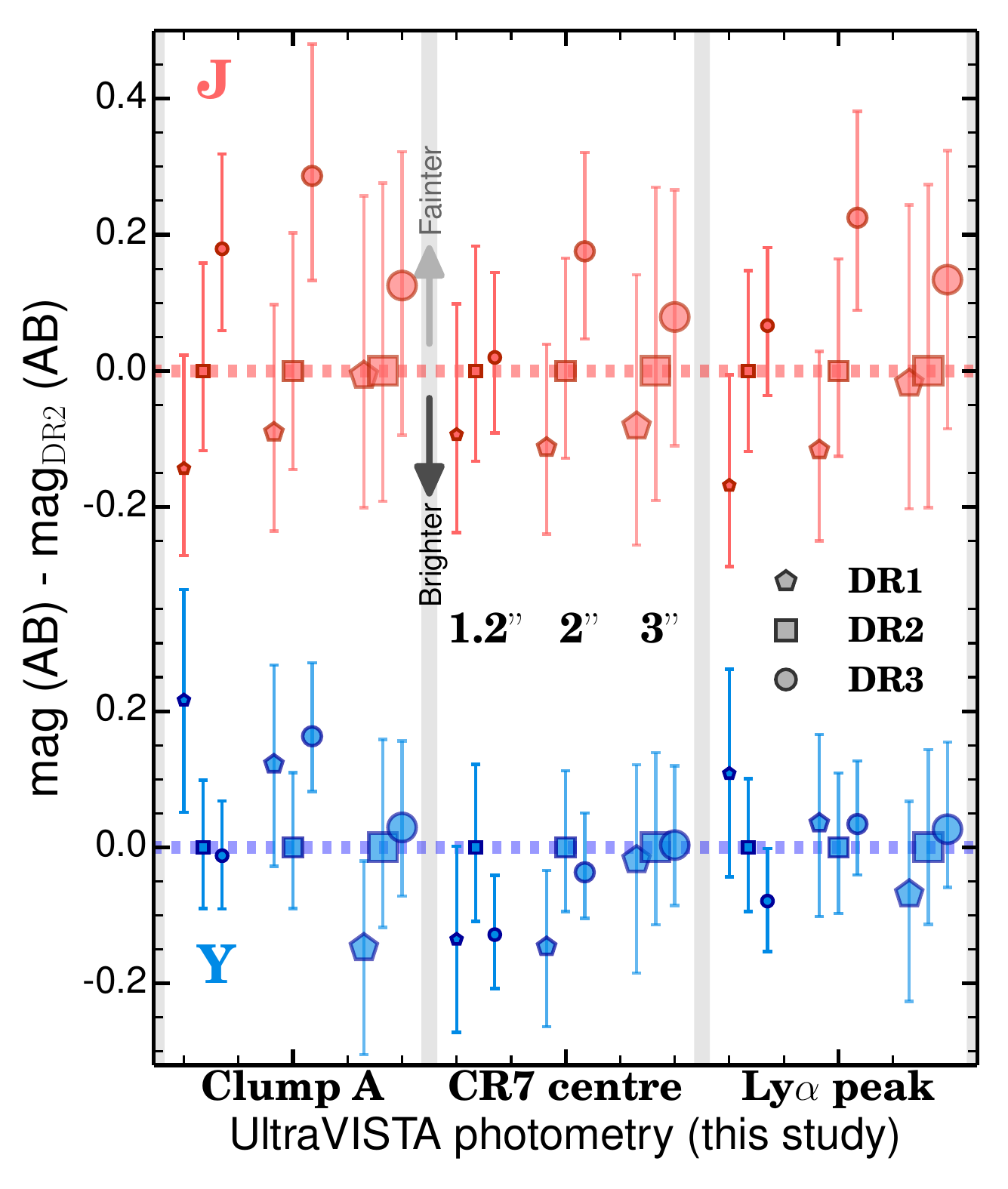}
\caption{{The difference in magnitude when compared to measurements of DR2 from the different UltraVISTA DRs. We measure magnitudes with apertures (diameter) of $1.2''$, $2.0''$ and $3.0''$ centred on clump A, on the centre of the 3 clumps and on the peak of Ly$\alpha$ emission for $Y$ and $J$ and compare them with the same measurement for DR2. We find no statistically significant variation from the different DRs, with only tentative dimming in the $J$ band from DR1 to DR2 and DR3.}}
\label{Variability_UVista}
\end{figure}

\section{Variability in {\it HST} photometric data}\label{var_HST}

{In order to measure or constrain any potential variability of CR7, as whole or in individual UV clumps, we use {\it HST} data taken on 2012-03-02, which was presented and explored in \citetalias{Sobral2015} ({\it HST} Program 12578), but we also use recent public data taken on 2017-03-14 ({\it HST} program 14596). For both {\it HST} programs, filters F110W and F160W were used. We first register a $30''\times30''$ cut-out of the four available stacks assuring that all sources within the image are fully aligned. We then use $ZP=26.6424,25.7551$ for F110W and F160W\footnote{Zero-points are found in: \url{http://www.stsci.edu/hst/wfc3/analysis/ir_phot_zpt}.}, respectively. We measure the flux and AB magnitudes in both filters for both dates, with apertures varying from 0.2 to 3$''$ in steps of 0.1, centred on the UV centroid of clumps A, B and C based on the stack of 2012 and 2017 data, and also centred on the rough centre of the full system (Figure \ref{XSHOOTER_angle}). In order to correct the aperture magnitudes we apply corrections for the missed flux of point sources, which vary from $\approx0.6-0.7$ for the smallest apertures to $\approx0.95$ for the largest\footnote{We use the corrections provided in: \url{http://www.stsci.edu/hst/wfc3/analysis/ir_ee}.}. In order to estimate the magnitude errors for a specific aperture and to measure a specific clump/location, we place 1,000 empty apertures throughout the image, avoiding bright sources (exploring a segmentation map produced with {\sc sextractor}) and compute the 16th and 84th percentiles, which we fold through to obtain magnitude errors. We also calculate the median of the flux measured in those 1,000 empty apertures and subtract it from the appropriate measurement, with the assumption that the median flux on locations without sources is a good proxy for the background at the location where we make the measurements.}

\section{Spectroscopy methodology: comparison with other studies}\label{Comparison_MC}

In order to evaluate and compare our methods and results for CR7 in the context of the discussions in this paper and e.g. \cite{Shibuya2017} we use X-SHOOTER data for recent studies. We explore other sources with detected emission-lines at $z\sim7-8$ that are publicly available. These are very helpful to compare the results from different statistical analysis and to also compare the reproducibility of results. We use recent very deep follow-up observations that have detected multiple lines \citep{Laporte_2017ALMA,Laporte2017} with X-SHOOTER targeting four different $z\sim6-8$ sources \citep[which have also been discovered or studied by other authors, e.g.][]{Stark2017,Smit2017}. We follow the procedure presented in \cite{Sobral2018b} and used in this paper. We focus on how well we recover the different rest-frame UV lines and how the S/N we measure compares with those reported in the literature.

For CR7 we have extracted the 1D spectra at the expected (centre) position of CR7 in the VIS arm by checking it matches with the rough peak of Ly$\alpha$ emission in the spatial direction. For the NIR arm, we extract the 1D along the central pixel for OB1 and OB2 where we do not find any emission line in the 2D, while for OB3 we extract $+3$ pixels away from the centre, extracting over $\pm1''$ in the VIS and NIR arms ($\pm8$ to $\pm4$ pixels depending on the arm). We follow the same methodology for the other X-SHOOTER spectra we study, taking care to extract over the signatures identified in the papers presenting the data and we extract over $\pm6$ spatial pixels in the VIS arm and $\pm3$ spatial pixels in the NIR arm to account for the fact that sources are typically more compact than CR7.

As for our main analysis of CR7, we start by using the errors provided by the pipeline reduction, but also investigate the S/N distribution across each X-SHOOTER arm. We find that the noise is typically over-estimated for our extractions based purely on the pipeline noise by factors of about 1.1 in the VIS arm and factors from 1.4-1.1 in the NIR arm \citep[see also e.g.][]{Zabl2015}. We re-measure the noise and check that the S/N of extracted spectra without any expected signal resemble Gaussian distributions. By using the pipeline noise directly we find that the S/N of empty regions is underestimated, but our final noise estimates yields a Gaussian S/N distribution for extractions consistent with no extragalactic signal.

%
%
\begin{figure}
\includegraphics[width=8.4cm]{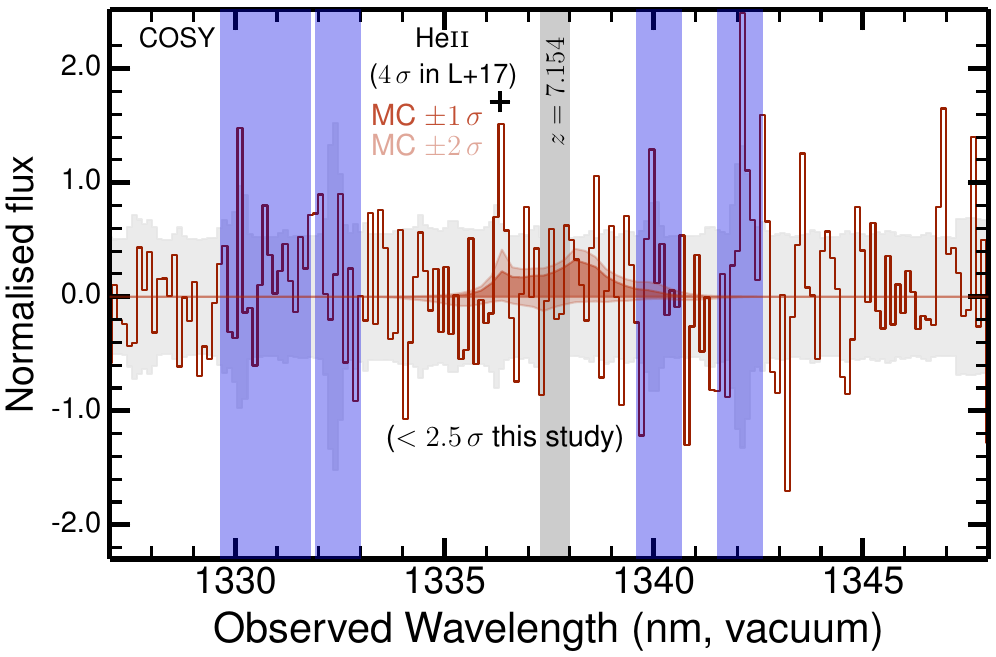}
\caption{Our analysis applied to COSY \citep[][]{Stark2017} and its potential He{\sc ii} line detection in X-SHOOTER data \citep[][]{Laporte2017}. We show the extracted 1D spectrum binned by one third of the resolution and also highlight the position of OH lines. Shaded regions show the $\pm1\,\sigma$ errors (grey) and the range of MC fits within 1 and 2\,$\sigma$, following Figure \ref{XSHOOTER_1D_MCMC_HeII}. We identify the signature identified as He{\sc ii} in our 1D and indicate it with a +. However, we find that such tentative signal corresponding to a redshift of $z\approx7.15$, reported to have a significance of $\approx4$\,$\sigma$ in \citet{Laporte2017} is below $2.5$\,$\sigma$ in our forward-modelling MC analysis.} 
\label{Comparison_Laporte_HeII}
\end{figure} 

Our re-analysis of data from the literature is able to recover spectra that resemble those in the literature. For Ly$\alpha$ emission, we agree with 3/4 detections, although we tend to find slightly lower S/N for those lines and also note that such Ly$\alpha$ lines (e.g. COSY) are actually very narrow. However, for other lines, out of 4 reported detections we only recover 2 lines at a S/N\,$>2.5$. This means that two of the lines reported to be at the $\approx4$\,$\sigma$ level in the literature, are found to be below $<2.5$\,$\sigma$ in our MC analysis. This is similar to the decreased significance between our study and \citetalias{Sobral2015} for He{\sc ii} in CR7, and it is likely a direct consequence of how the noise/significance is measured, along with effects of smoothing/binning. We show examples in Figures \ref{Comparison_Laporte_HeII} and \ref{Comparison_Laporte_Lya}. Here we list the results for the sources investigated:

\begin{itemize}

 \item {\bf COSY \citep[][]{Laporte2017}}: We confirm Ly$\alpha$ at $z=7.1542^{+0.0007}_{-0.0009}$, in full agreement with what had been found by \cite{Stark2017} with a MOSFIRE spectrum and what is also concluded in \cite{Laporte2017}. However, we note that contrarily to the discussion presented in \cite{Laporte2017}, we find that COSY's Ly$\alpha$ line is not unusually broad, but rather relatively narrow for a Ly$\alpha$ line (see also Figure \ref{Comparison_Laporte_Lya}). We find that its FWHM (deconvolved for resolution) is $312^{+27}_{-32}$\,km\,s$^{-1}$, and thus consistent with being as narrow as the Ly$\alpha$ line from CR7. COSY's Ly$\alpha$ line is very narrow given how bright in the rest-frame UV this source is, but this seems to be a general feature of Ly$\alpha$ emitters in the epoch of re-ionisation \citep[see][]{Matthee2017,Sobral2018b}. Apart from Ly$\alpha$, we find no other emission line in COSY in the X-SHOOTER spectra above 2.5\,$\sigma$. The reported detections of N{\sc v} and He{\sc ii} at $\approx4$\,$\sigma$ are all below $2.5$\,$\sigma$ in our analysis. Specifically, we find that the reported N{\sc v} detection is consistent with the noise level and the proximity to a strong OH line. For He{\sc ii}, while there is a tentative signal (see Figure \ref{Comparison_Laporte_HeII}), the peak of the signal is too narrow, while the full signal is not significant. The potential He{\sc ii} signal for COSY, if measured manually (as our automated analysis does not detect it), would be consistent with a very low FWHM of $\approx$50\,km\,s$^{-1}$, below the resolution. We note that \cite{Laporte2017} also presents a MOSFIRE spectrum that seems find He{\sc ii} for COSY, but here we focus on X-SHOOTER and we aim to only report our findings in our framework.
 
%
%
\begin{figure}
\includegraphics[width=8.4cm]{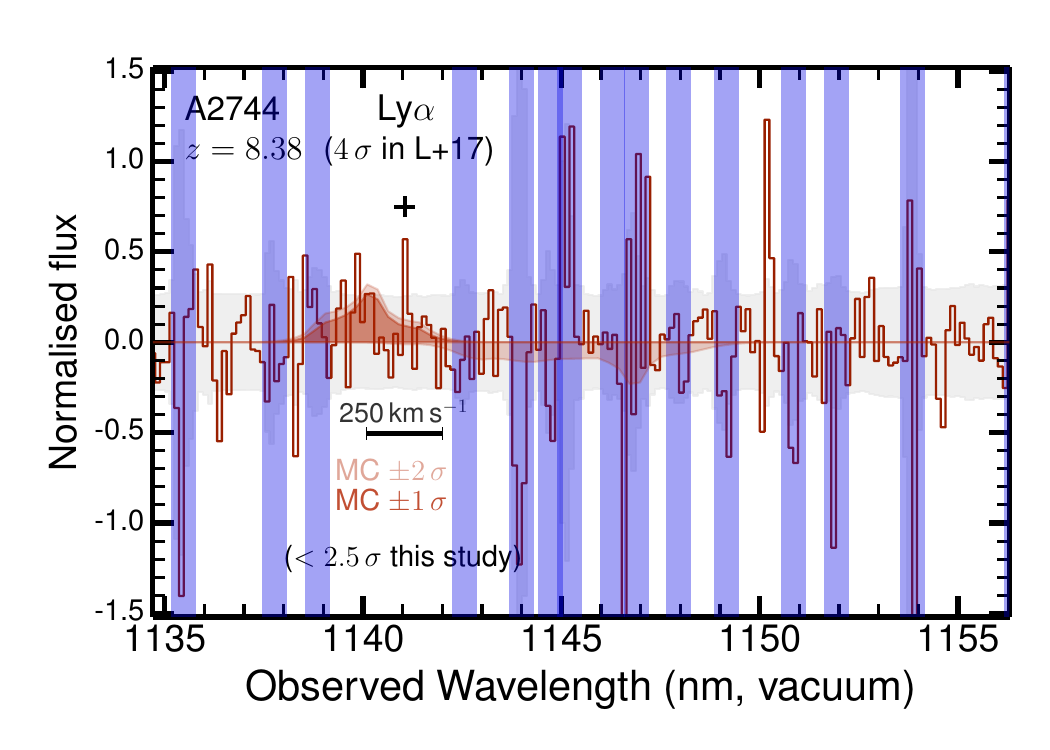}
\caption{Our analysis applied to A2744 and its potential Ly$\alpha$ line detection in X-SHOOTER data \citep[][]{Laporte_2017ALMA}. We show the extracted 1D spectrum binned by one third of the resolution and also highlight the position of OH lines. We recover and indicate the signal interpreted as Ly$\alpha$ in \citet{Laporte_2017ALMA} with a +. The potential line is reported to have a significance of $4$\,$\sigma$ in \citet{Laporte_2017ALMA}, but our methodology implies that any signal is below a significance of $2.5$\,$\sigma$. Furthermore, we also indicate the expected width of a very narrow Ly$\alpha$ line with a FWHM of 250\,km\,s$^{-1}$, which is significantly broader than the single spectral element identified as an emission line in \citet{Laporte_2017ALMA}.}
\label{Comparison_Laporte_Lya}
\end{figure}

 \item {\bf COSz1 \citep[][]{Laporte2017}}: We recover the C{\sc iii}]1909 emission line above 3\,$\sigma$ (3.4\,$\sigma$). The detection of the line is consistent with \cite{Laporte2017}, despite our detection at lower significance (3.4\,$\sigma$ instead of 4\,$\sigma$), but the difference is small. We also note that for C{\sc iii}] \cite{Laporte2017} seem to have binned the spectra at least to a fraction of the resolution; while that is not always the case for the other lines, and particularly not the case for the lines in other sources which we find to be below 2.5\,$\sigma$ and below the resolution of the instrument. Apart from C{\sc iii}], no other emission line is found in our analysis above 2.5\,$\sigma$, which is in agreement with \cite{Laporte2017}.

 \item {\bf COSz2 \citep[][]{Laporte2017}}: We confirm Ly$\alpha$ and no other emission lines from this source above $2.5$\,$\sigma$ in our analysis, in full agreement with \cite{Laporte2017}. Due to the overlap with a strong OH line we find that the Ly$\alpha$ line is detected at just below $3$\,$\sigma$; \cite{Laporte2017} reports its significance as $\approx3$\,$\sigma$, and thus we conclude there is good agreement. Furthermore, we identify a significant emission line at $\approx1552$\,nm, also in full agreement with \cite{Laporte2017}, which is argued in that paper to be from a source at $z\sim2$ and to potentially be [O{\sc iii}]5007, since it does not match any potential line for the redshift of COSz2. 
 
\item {\bf A2744 \citep[][]{Laporte_2017ALMA}}: We find no emission lines detected above $2.5$\,$\sigma$ on the entire X-SHOOTER spectra. In particular, while we can tentatively identify the signal of the reported 4\,$\sigma$ detection of Ly$\alpha$ in the 2D spectrum and explicitly extract the spectrum centred on that, our analysis reveals it is not statistically significant; see Figure \ref{Comparison_Laporte_Lya}. We find that the signal in \cite{Laporte_2017ALMA} comes from too few pixels and is below the resolution, implying a FWHM of $\approx$\,20\,km\,s$^{-1}$. Given that the resolution, measured with nearby sky lines on the spectrum, is close to 60\,km\,s$^{-1}$, a potential emission line with a FWHM of $\approx$\,20\,km\,s$^{-1}$ is below the resolution. This means that this line would have a FWHM about 3 times lower than the OH lines; this is something more typical of noise and/or artefacts, as any real line will have at least a FWHM equal to the resolution, even if intrinsically it is even narrower. We therefore conclude that with our conservative statistical analysis that we apply to CR7, what is measured as a 4\,$\sigma$ Ly$\alpha$ line in \cite{Laporte_2017ALMA} for A2744 is consistent with noise or an artefact and it is below 2.5\,$\sigma$ in our framework, and thus we would report it as undetected. In Figure \ref{Comparison_Laporte_Lya} we also show how a very narrow Ly$\alpha$ line with a FWHM of 250\,km\,s$^{-1}$ should look like in the spectrum, significantly broader than the potential line indicated with +.

\end{itemize}

\bsp	
\label{lastpage}
\end{document}